\documentclass[twoside,11pt]{article}

\usepackage[accepted]{melba}

\usepackage[utf8]{inputenc}

\usepackage{amsfonts}
\usepackage{amsmath}
\usepackage{mathrsfs}
\usepackage{amsthm}

\usepackage{algorithm}
\usepackage{algorithmic}

\usepackage{subfig}
\usepackage{xcolor}

\usepackage[toc,page]{appendix}

\newcommand{\x}{\vec{\mathbf{x}} } 

\newcommand{\y}{\vec{\mathbf{y}}} 

\newcommand{\Src}{S} 
\newcommand{\Tgt}{T} 
\newcommand{\Srcn}{{S_n}}

\newcommand{\Det}[1]{\left| #1 \right|}

\newcommand{\USd}{\mathbb{S}^{d-1}}
\newcommand{\RxSd}{\Rd \times \USd}

\newcommand{\dps}{\displaystyle}
\newcommand{\om}{\omega}

\newcommand{\R}{\mathbb{R}}
\newcommand{\norm}[1]{\left\lVert#1\right\rVert}
\newcommand{\Rd}{\mathbb{R}^d}

\newcommand{\hyperfootnote}[1][]{\def\ArgI\hyperfootnoteRelay}
\newcommand\restr[2]{\ensuremath{{#1}_{|#2}}}

\newtheorem{Proposition}{Proposition} 
\newtheorem{Definition}{Definition} 

\melbaheading{2022:006}{https://www.melba-journal.org/papers/2022:006.html}{2022}{1-29}{12/2021}{04/2022}{Antonsanti, Benseghir, Jugnon, Ghosn, Chassat, Kaltenmark and Glaunès}{Information Processing in Medical Imaging (IPMI) 2021}{Aasa Feragen, Stefan Sommer, Julia Schnabel, Mads Nielsen}

\ShortHeadings{How to Register a Live onto a Liver ?}{Antonsanti et al.}
\firstpageno{1}

\title{\vspace{-1em}How to Register a Live onto a Liver ? \\ Partial Matching in the Space of Varifolds}

\author{\name Pierre-Louis Antonsanti \email pierrelouis.antonsanti@ge.com \\  
	\addr{GE Healthcare, Buc 78530, France}
	\AND
	\name Thomas Benseghir \email thomas.benseghir@ge.com \\
	\addr{GE Healthcare, Buc 78530, France}
	\AND
	\name Vincent Jugnon \email vincent.jugnon@ge.com \\
	\addr{GE Healthcare, Buc 78530, France}
	\AND
	\name Mario Ghosn \email ghosnm@mskcc.org \\
	\addr{Memorial Sloan Kettering Cancer Center, New York, NY USA}
	\AND
	\name Perrine Chassat \email perrine.chassat@univ-evry.fr \\
	\addr{Laboratoire de Math\'ematiques et Mod\'elisation d'\'Evry, \'Evry 91037, France}
	\AND
	\name Ir\`ene Kaltenmark \email irene.kaltenmark@u-paris.fr \\
	\addr{MAP5, Universit\'e de Paris, 45 Rue des Saints P\`eres, Paris 75006, France}
	\AND
	\name Joan Glaun\`es \email alexis.glaunes@u-paris.fr \\
	\addr{MAP5, Universit\'e de Paris, 45 Rue des Saints P\`eres, Paris 75006, France}
}

\begin{document}

\vspace{-2em}
\maketitle
\vspace{-3em}
\begin{abstract}
Partial shapes correspondences is a problem that often occurs in computer vision (occlusion, evolution in time...).
In medical imaging, data may come from different modalities and be acquired under different conditions which leads to variations in shapes and topologies.
In this paper we use an asymmetric data dissimilarity term applicable to various geometric shapes like sets of curves or surfaces, assessing the embedding of a shape into another one without relying on correspondences. 
It is designed as a data attachment for the Large Deformation Diffeomorphic Metric Mapping (LDDMM) framework, allowing to compute a meaningful deformation of one shape onto a subset of the other.
We refine it in order to control the resulting non-rigid deformations and provide consistent deformations of the shapes along with their ambient space. 
We show that partial matching can be used for robust multi-modal liver registration between a Computed Tomography (CT) volume and a Cone Beam Computed Tomography (CBCT) volume. 
The 3D imaging of the patient CBCT at point of care that we call \textit{live} is truncated while the CT pre-intervention provides a full visualization of the \textit{liver}. 
The proposed method allows the truncated surfaces from CBCT to be aligned non-rigidly, yet realistically, with surfaces from CT with an average distance of $2.6mm(\pm 2.2)$. The generated deformations extend consistently to the liver volume, and are evaluated on points of interest for the physicians, with an average distance of $5.8mm (\pm 2.7)$ for vessels bifurcations and $5.13mm (\pm 2.5)$ for tumors landmarks. 
Such multi-modality volumes registrations would help the physicians in the perspective of navigating their tools in the patient's anatomy to locate structures that are hardly visible in the CBCT used during their procedures.
Our code is available at~\url{https://github.com/plantonsanti/PartialMatchingVarifolds}.
\end{abstract}




\section{Introduction}

In medical imaging, the problem of registering images has been tackled by numerous authors \citep{Sotiras2013} by registering directly the images, most of the time assuming that both images contain the entire object of interest.
When dealing with similar images (e.g. same acquisition modality of a given patient), techniques driven by a distance between voxel intensities (called similarity) usually perform well \citep{Bauer2021}.
When more variability in terms of intensity arises, other studies propose feature based methods that match structures of interest such as surfaces, curves or even patches \citep{Jiang2021}, showing better robustness than their intensity based counterparts.

The problem of finding correspondences between these structures has numerous applications such as pattern recognition \citep{Bronstein2006, Bronstein2009, Kaick2013}, annotation (\cite{Benseghir2013}, \cite{Feragen2015}), and reconstruction \citep{Halimi2020}.
In particular, in the field of medical imaging, matching an atlas and a patient's anatomy \citep{Feragen2015},
or comparing exams of the same patient acquired with different imaging techniques \citep{Bashiri2018}, provide critical information to physicians for both planning and decision making.
In practice, it often happens that only part of an object is visible in one of the two modalities (e.g. disease, surgery, multi-modality imaging, noise). 
When only partial correspondence between the images are present, handling missing structures is often done manually or heuristically.

Cone Beam Computed Tomography (CBCT), for instance, is used during image guided procedures -live- to improve navigation and guidance. Yet, the imaged organs are generally larger than the field of view, while the whole organ such as the liver can be acquired in Computed Tomography (CT) scanners during the disease diagnostic phase. 
\begin{figure*}[!ht]
\centering
   \subfloat[\label{fig:CBCT}]{%
      \includegraphics[clip, width=0.4\textwidth]{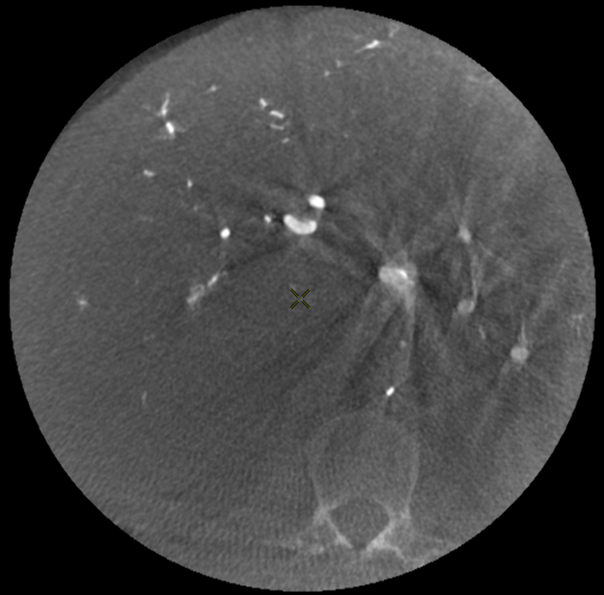}}
\hspace{.5cm}
   \subfloat[\label{fig:CT}]{%
      \includegraphics[clip, width=0.4\textwidth]{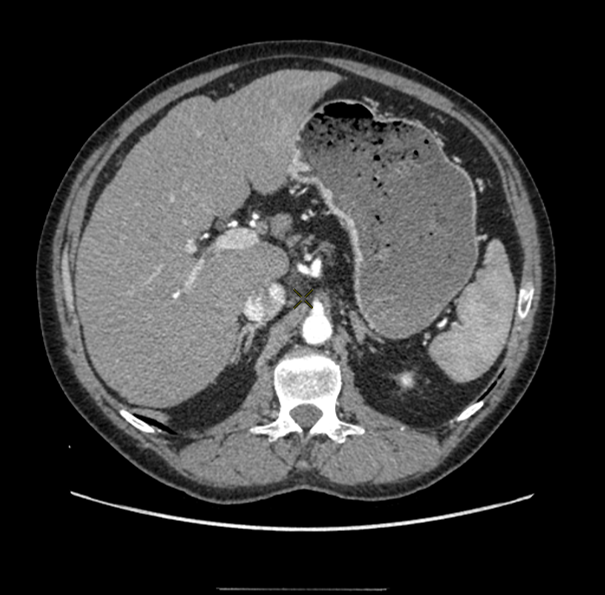}}

\caption{\label{fig:illustration_modalities} Examples of slices from a CBCT volume (a) and CT volume (b) from a same patient visualized with a 3mm Maximum Intensity Projection (MIP). The liver is only partially visible in CBCT.}
\end{figure*}

In order to make the best of the different images both in terms of field of view and image characteristics (see Fig.~\ref{fig:illustration_modalities}), one can find a partial matching between them and the feature-based approaches seem suited to such matching problems. 
This work is focused on the registration of shapes where only part of these structures can be matched and its application to the registration of liver surfaces extracted from CT and CBCT to guide multi-modal volumes registration. 
In particular we apply the registration to the whole volumes, and must control the deformations to generate anatomically relevant ones. 
We study the clinical application of the proposed partial matching term for The present paper is an extension of \cite{Antonsanti2021} from the 2021 Information Processing in Medical Imaging conference in which we introduced the partial matching in the space of varifolds.

\subsection{Previous works}
The problem of matching shapes has been widely addressed in the literature in the past decades \citep{Kaick2011}. 
In the specific case of partial matching, one can find two main approaches to such problem: 
either by finding correspondences, sparse or dense, between structures from descriptors that are invariant to different transformations \citep{Aiger2008, Rodola2017, Halimi2019}, or by looking for a deformation aligning the shapes with respect to a given metric \citep{Aiger2008, Benseghir2013, Zhao2015, Feragen2015, Halimi2020}.

The early works on partial shape correspondence as reviewed in \cite{Kaick2011} rely on {\bf correspondences} between points computed from geometric descriptors extracted from an isotropic local region around the selected points. 
The method is refined in \cite{Kaick2013} by selecting pairs of points to better fit the local geometry using bilateral map.
The features extracted can also be invariant to different transformation, as in \cite{Rodola2013} where the descriptors extracted are scale invariant. 
Such sparse correspondences are naturally adapted to partial matching, yet they cannot take the whole shapes into account in the matching process. 

Using a different approach, functional maps were introduced in \cite{Ovsjanikov2013} allowing dense correspondences between shapes by transferring the problem to linear functions between spaces of functions defined over the shapes. 
In \cite{Rodola2017} the non-rigid partial shape correspondence is based on the Laplace-Beltrami eigenfunctions used as prior in the spectral representation of the shapes. 
Recently in \cite{Halimi2019}, such functional map models were adapted in a deep unsupervised framework by finding correspondences minimizing the distortion between shapes. Such methods are yet limited to surface correspondences. 

The second kind of methods relies on the {\bf deformations} that can be generated to align the shapes with each other. 
It usually involves minimizing a function called data attachment that quantifies the alignment error between the shapes. 
A deformation cost is sometimes added to regularize these deformations. 
The sparse correspondences being naturally suited to partial matching, they are notably used in Iterative Closest Point (ICP) methods and their derivatives to guide a registration of one shape onto the other. 
In \cite{Aiger2008} a regularized version of the ICP selects the sets of four co-planar points in the points cloud. 
In \cite{Benseghir2013} the ICP is adapted to the specific case of vascular trees and compute curves' correspondences through an Iterative Closest Curve method. 
Working on trees of 3D curves as well, \cite{Feragen2015} hierarchically selects the overall curves correspondences minimizing the tree space geodesic distance between the trees. 
This latter method, although specific to tree-structures, allows topological changes in the deformation. 

On the other hand some authors compute the deformation guided by a dense data attachment term. In \cite{Bronstein2006} isometry-invariant minimum distortion deformations are applied to 2D Riemannian manifolds thanks to a multiscale framework using partial matching derived from Gromov's theory. This work is extended in \cite{Bronstein2009}  by finding the optimal trade-off between partial shape matching and similarity between these shapes embedded in the same metric space.
Recently in \cite{Halimi2020} a partial correspondence is performed through a non-rigid alignment of one shape and its partial scan seen as points clouds embedded in the same representation space. 
The non-rigid alignment is done with a Siamese architecture network. 
This approach seems promising and is part of a completion framework, however it requires a huge amount of data to train the network. 

Interestingly, the regularization cost can be seen as a distance between shapes itself \citep{Feragen2015} by quantifying the deformation amount necessary to register one shape onto the other. 
This provides a complementary tool to the metrics used to quantify the shapes dissimilarities. 

A well established and versatile framework to compute meaningful deformations is the Large Deformation Diffeomorphic Metric Mapping (LDDMM) \citep{Trouve1995, Christensen1996, Beg2005}.
It allows to see the difference between shapes through the optimal deformation to register a source shape $\Src$ onto a target shape $\Tgt$. 
However, the data attachment metrics proposed so far aim to compare the source and target shapes in their entirety \citep{Charon2020}, or look for explicit correspondences between subparts of these shapes \citep{Feydy2017}.
In \cite{Kaltenmark2018} the proposed growth model introduces a first notion of partial matching incorporated to the LDDMM framework. More recently, the bijectivity constraint is circumvented by the introduction of weighted shapes: in \cite{Hsieh2021, Sukurdeep2021}, authors optimize a mask defined on the shapes to exclude some subparts of these shapes (source or target).
In \cite{Antonsanti2021} a partial matching data fidelity term was introduced in the space of varifolds. We use it in this paper and extend its formulation to better control the non-rigid deformations. We then apply the overall method to guide multi-modal volumes registration.

\subsection{Organization of the Paper}

In the first part we will shortly review the main elements of the representation of shapes in the space of varifolds \citep{Charon2013} and the expression of the partial matching we use in this space, introduced in \cite{Antonsanti2021}. In particular, we recall how it can be used with LDDMMs to compute diffeomorphic registration of a source shape to a subset of a target one. We will also discuss the use of the partial matching with a rigid registration. We then provide more details on the regularization term used for the application to the registration of truncated liver surfaces to limit the shrinkage effect that can occur when using partial matching with LDDMM.

The second section of this paper is dedicated to its clinical application to multi-modality volume registration based on the partial matching of liver surfaces. 
In this application we register liver surfaces obtained from segmentations of pre-interventional CT volumes intended for diagnostic purposes, and per-interventional CBCT volumes in which the livers are wider than the field of view, leading to truncation of the surfaces in the latter modality.


\section{Partial Matching in the space of varifolds}\label{sec:partial_matching}

In this section we quickly review the methodology introduced in \cite{Antonsanti2021}. We refer to this article for all details.

We are interested in the problem of finding an optimal deformation to register a source shape $\Src$ onto an unknown subset of a target shape $\Tgt$, where $\Src, \Tgt$ are assumed to be compact $m$-rectifiable subsets of $\Rd$ with either $m=1$ (curves) or $m=d-1$ (hypersurfaces). 

\subsection{Shape registration via oriented varifolds}

The varifold method for shape matching was first introduced in \cite{Charon2013}. In this section we use the notations and concepts of the more general oriented varifold framework, developed in \cite{Kaltenmark2017}. This framework associates with any shape $\Src$ a {\it representer} function, defined on $\RxSd$, as follows:
$$\omega_\Src(y,\tau)=\int_\Src k_e(y,x)k_t(\tau,\tau_x\Src)dx \,,$$
where $\tau_x$ denotes the unit tangent vector --for curves-- or unit normal vector --for hypersurfaces-- at point $x$ on the shape $\Src$, and $k_e, k_t$ are predefined fixed kernels functions over $\Rd$ and $\USd$. 
In practice, we use
$k_e:{(\Rd)}^2\rightarrow\R$ a gaussian kernel $k_e(x,y)=e^{-\|x-y\|^2/\sigma_W^2}$, where $\sigma_W$ is a scale parameter, and for $k_t:{(\USd)}^2\rightarrow\R$ the kernel $k_t(u,v) = e^{\langle\,u,v\rangle_{\Rd}}$.
Mathematically, the product kernel $k_e\otimes k_t$ defines a Reproducing Kernel Hilbert Space (RKHS) structure $W$, and $\omega_\Src$ corresponds to the Riesz representer of the {\it oriented varifold} $\mu_\Src\in W'$, associated with $\Src$, and $W'$ the space of continuous linear forms on $W$.

The shape matching dissimilarity between $\Src$ and $\Tgt$ is then the squared $W$-norm between their representers, which is also the squared $W'$ (dual) norm between their varifolds:

\begin{eqnarray*}
d_{W'}(\Src,\Tgt)^2 &=& \Vert \omega_{\Src} - \omega_{\Tgt} \Vert_{W}^2 = \Vert \mu_{\Src} - \mu_{\Tgt} \Vert_{W'}^2 \\&=& \langle\,\mu_{\Src},\mu_{\Src}\rangle_{W'}-2\langle\,\mu_{\Src},\mu_{\Tgt}\rangle_{W'}+\langle\,\mu_{\Tgt},\mu_{\Tgt}\rangle_{W'}\,,
\end{eqnarray*}

where the scalar product has an explicit form using the kernels:
\begin{eqnarray*}
\langle\,\mu_\Src,\mu_\Tgt\rangle_{W'}
=\langle\,\omega_\Src,\omega_\Tgt\rangle_{W}
=\int_\Src\int_\Tgt k_e(x,y)k_t(\tau_x\Src,\tau_{y}\Tgt)dx\,dy\,.
\end{eqnarray*}

\subsection{Definition of the Partial Matching Dissimilarity}

 
The proposed partial dissimilarity term is designed to look locally at the inclusion of the deformed source shape into the target one, and to compensate (with the $max(.,.)$ function) for the imbalance of weights between the simple shape and the more complex one with a normalization. The following term tends to 0 when the deformed shape is included in the target one (for more details and discussion, see \cite{Antonsanti2021}). 
 
\begin{Definition}\label{def:partial_normalized_dissimilarity} 
Let $g : \mathbb{R} \mapsto \mathbb{R}$ defined as $g(s) = (\max(0,s))^2$, and denote for $x,x' \in S$, $\x = (x,\tau_x S)$, $\om_S(\x)=\omega_S(x,\tau_x\Src)$ and $k(\x,\x') = k_e(x,x')k_t(\tau_x\Src,\tau_{x'}\Src)$. We define the {\bf normalized partial matching dissimilarity} as follows: 

\begin{eqnarray}
\underline{\Delta}(\Src,\Tgt) &=& 
 \int_{\Src}g\left(\om_S(\x)- \int_\Tgt     {\min}_\epsilon \left( 1, \dfrac{\om_{\Src}(\x)}{\om_\Tgt(\y)} \right) k(\x,\y) dy \right) dx
\end{eqnarray}
where $\min_\epsilon(1,s)=\frac{s+1-\sqrt{\epsilon+(s-1)^2}}{2}$ with $\epsilon>0$ small, is used as a smooth approximation of the $\min(1,\cdot)$ function.
\end{Definition}

\paragraph{Discrete formulation}\label{sub:discrete}
The discrete version of the partial matching dissimilarity can be derived very straightforwardly, following the same discrete setting described in \cite{Charon2020}
for varifold matching. 
We are working with surfaces, seen as triangular meshes with vertices $q_1,...,q_K$. Each triangle $f_i,\: i\in[1,..,N_{\Src}]$ with the vertices $(q_i^1, q_i^2, q_i^3)$ of the shape $\Src$ is associated to the center $c_i = \frac{x_i^1 + x_i^2 + x_i^3}{3}$ and to the normal vector $\eta_{x_i}\Src = (1/2)*(q_i^2-q_i^1)\times(q_i^3-q_i^1)$. The unit normal vector is then $ \tau_{x_i}\Src = \frac{\eta_{x_i}\Src}{\norm{\eta_{x_i}\Src}}$.
We define similarly the centers $y_l$ of the target shape $\Tgt$ and their associated normal vectors $\eta_{y_l}\Tgt$ and unit normal vectors $\tau_{y_l}\Tgt$.

The discrete normalized partial matching term is then written as follows: 
\begin{eqnarray}
\underline{\Delta}(\Src,\Tgt) &= \sum\limits_{i=1}^{N_\Src}  g\left( \om_{\Src}(\x_i) - \sum\limits_{l=1}^{N_\Tgt}{\min}_\epsilon \left( 1, \dfrac{\om_{\Src}(\x_i)}{\om_\Tgt(\y_l)}\right) k(\x_i,\y_l)\right)
\end{eqnarray}\label{eq:discrete_pm}

with $ \om_{\Src}(\x_i) =  \sum\limits_{j=1}^{N_\Src}  k_e(x_i,x_j)k_t(\tau_{x_i}\Src,\tau_{x_j}\Src) \norm{\eta_{x_i}\Src} \norm{\eta_{x_j}\Src} $.\\

\subsection{Use in the LDDMM setting}\label{sec:LDDMM}
In the LDDMM framework \cite{Beg2005}, the partial matching problem consists in minimizing $$ J(v) = \lambda \int_0^1 \Vert v_t \Vert_V^2dt + \underline{\Delta}({\phi_1^v(\Src)},\Tgt)\,,$$ 
where $\phi_1^v$ is the flow of the time-dependent square integrable velocity fields $t \in [0,1] \mapsto v_t$, and $\|\cdot\|_V$ is the regularization Hilbert norm over vector fields.

The LDDMM registration procedure is numerically solved via a geodesic shooting algorithm introduced by \cite{Miller06}, optimizing on a set of initial momentum vectors located at the discretization points of the source shape.
Implementations of the dissimilarity terms are available on our github repository \footnote{\url{https://github.com/plantonsanti/PartialMatchingVarifolds}}.

\subsection{Use with Rigid Deformations}\label{sec:rigid}

We can define a second minimization problem for a rigid registration, by minimizing the function:
$$ J_{rig}(r) = \underline{\Delta}({r(\Src)},\Tgt)\,,$$
with $r$ a rigid deformation composed of a translation and a rotation. 

For any rigid deformation $\om_{r(\Src)}(r(\x)) = \om_{\Src}(\x)$, $\forall x \in \Src$, and we can thus write in the discrete setting:
\begin{eqnarray*}
\underline{\Delta}(r(\Src),\Tgt) &= \sum\limits_{i=1}^{N_\Src}  g\left( \om_{\Src}(\x_i) - \sum\limits_{l=1}^{N_\Tgt}{\min}_\epsilon \left( 1, \dfrac{\om_{\Src}(\x_i)}{\om_\Tgt(\y_l)}\right) k(r(\x_i),\y_l)\right)
\end{eqnarray*}\label{eq:rigid_solution}
We have that $J_{rig}$ is a composition of continuous functions and that : $k(r(\x_i),\y_l) \longrightarrow 0$ when the translation goes to infinity from the construction of the kernel $k$. We can deduce that for all rigid  deformation we have ${\min}_\epsilon \left( 1, \dfrac{\om_{\Src}(\x_i)}{\om_\Tgt(\y_l)}\right) k(r(\x_i),\y_l) > 0$ and $ J_{rig}(r) < \sum\limits_{i=1}^{N_\Src}  g\left( \om_{\Src}(\x_i)\right) $.\\

$J_{rig}$ is continuous and bounded over the space of finite dimension of rotations and translations, the minimization problem has a solution.

\subsection{Additional a priori}

It is known from experiments that the deformations can lead to abnormal shrinkage or stretching of the deformed shapes. 
This phenomenon comes from two things combined: first, we introduce an attachment term to the data favoring the inclusion of a deformed object in a target, thus an asymmetric term. In the case of non-rigid deformations, there is a multitude of local minima, and this must be controlled with a regularization of the deformations. Second, in the regularization of the LDDMM model (\ref{sec:LDDMM}), the deformations tend to shrink the objects along the geodesics, so it is possible that the diffeomorphisms create a shrinkage of the non-realistic source shape.
In order to limit this we can add a regularization term in the function $J$ to minimize.

\subsubsection{Definition}

The purpose of this regularization term is to prevent the deformations from shrinking or stretching the source shape. We chose to address it in the shape space of Varifolds as well, by controlling the norm of the deformed shape:

\begin{eqnarray}
R_{global}(\Src,\Phi(\Src)) = \left(\Vert \omega_{\Src}\Vert^2_{W} - \Vert \omega_{\Phi(\Src)} \Vert^2_{W} \right)^2 \quad\mbox{(\textbf{Global})}
\end{eqnarray}\label{eq:global_regul}

Interestingly this term can be written with the area formula:
\begin{align*}
\begin{split}
R_{global}(\Src,\Phi(\Src)) &=  \left(\int_\Src \om_\Src(\x) dx - \int_{\Phi(\Src)} \om_{\Phi(\Src)}(\y)dy \right)^2 \\
                            &= \left(\int_\Src \om_\Src(\x) - \om_{\Phi(\Src)}(\Phi(\x)) \Det{\restr{d_x\phi}{\tau_x S}} dx \right)^2 
\end{split}
\end{align*}

\begin{figure}[ht!]
\begin{minipage}{.98\linewidth}
\centering
   \subfloat[Source and Target \label{fig:regul_source}]{%
      \includegraphics[clip, width=0.3\textwidth, trim=4cm 3cm 4cm 3cm]{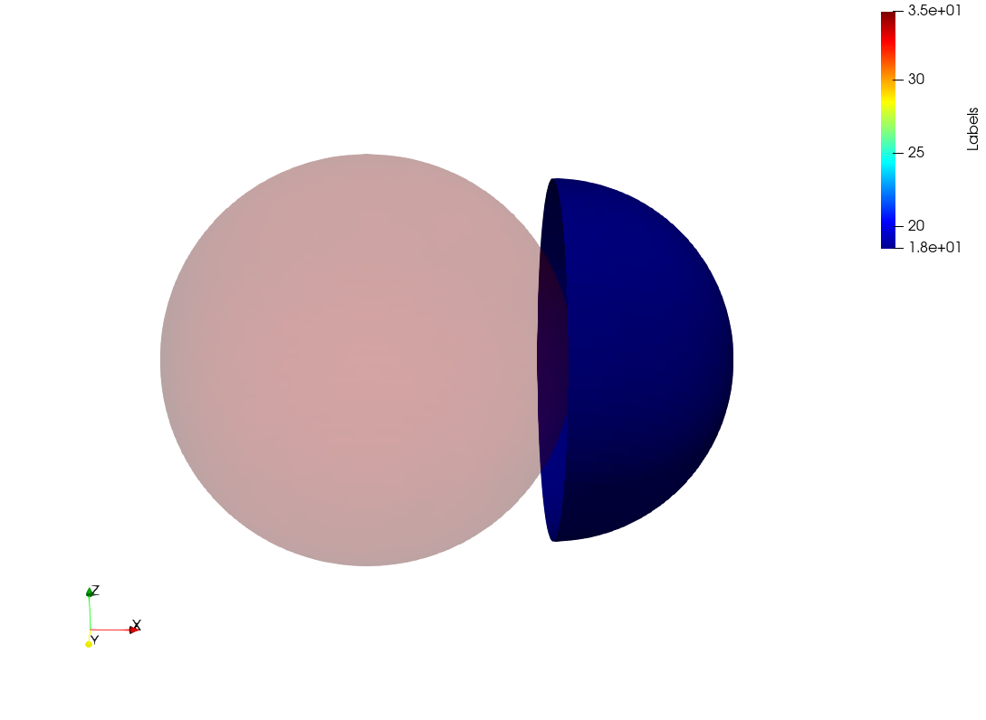}}
\hspace{0.05\textwidth}
\centering
\centering
   \subfloat[Partial Varifold \\No Regularization\label{fig:pv_no_regul}]{%
      \includegraphics[clip, width=0.3\textwidth, trim=4cm 3cm 4cm 3cm]{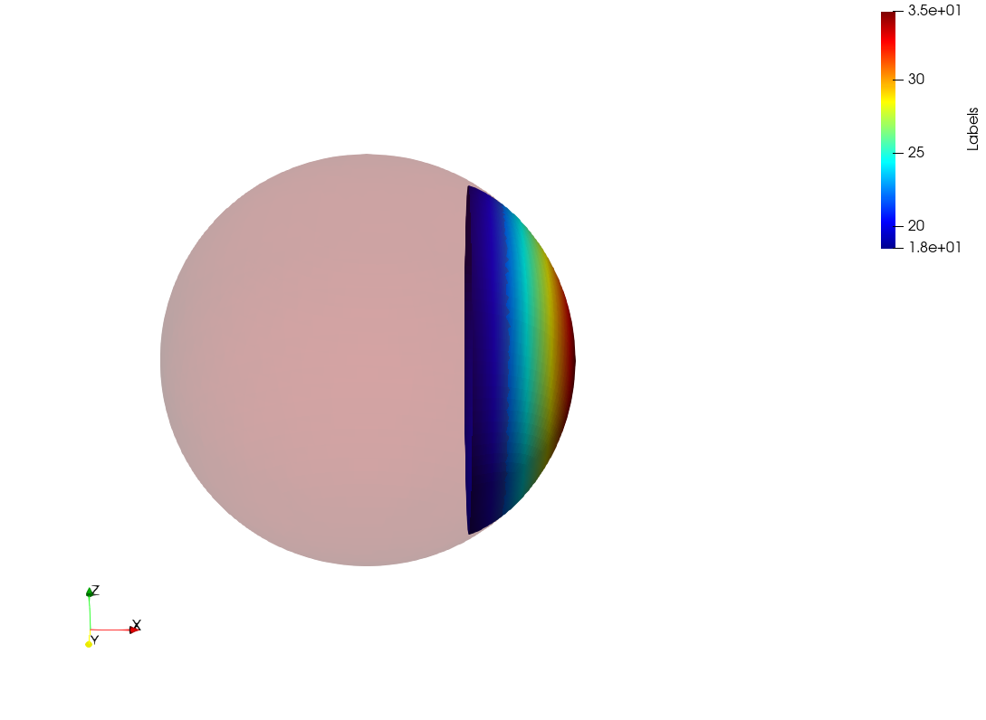}}
\centering
   \subfloat[\empty \label{fig:regul_colorbar}]{%
      \includegraphics[clip, width=0.1\textwidth, trim=22cm 12cm 0.6cm 0cm]{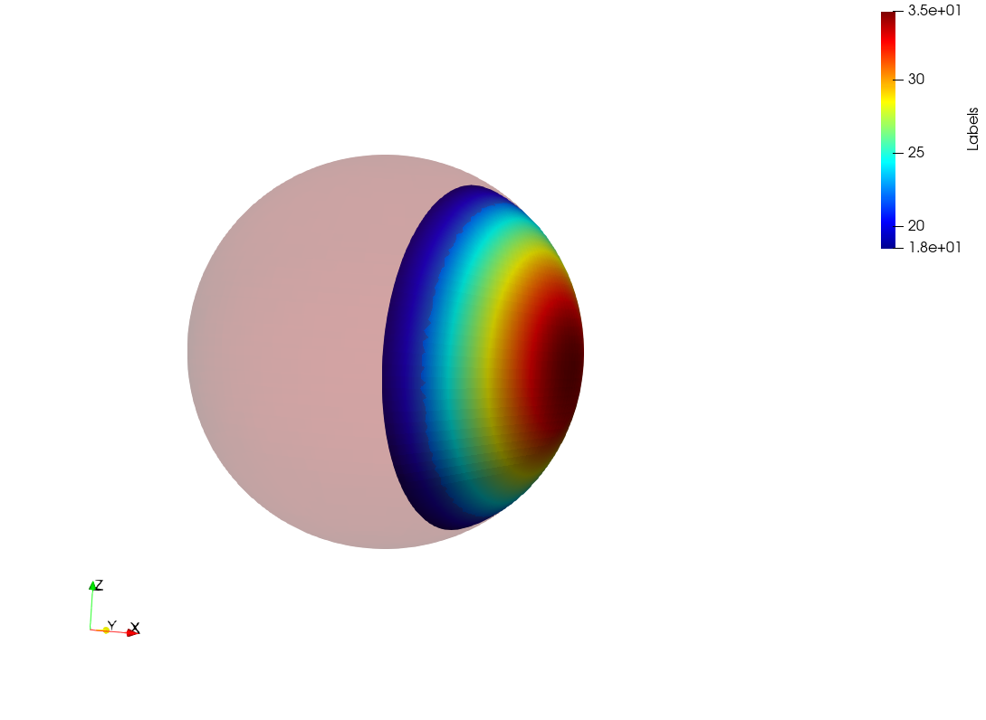}}
\centering
\par
\vspace{0.05\textwidth}
   \subfloat[Partial Varifold \\ Global Regularization \label{fig:pv_regul_global}]{%
      \includegraphics[clip, width=0.3\textwidth, trim=4cm 3cm 4cm 3cm]{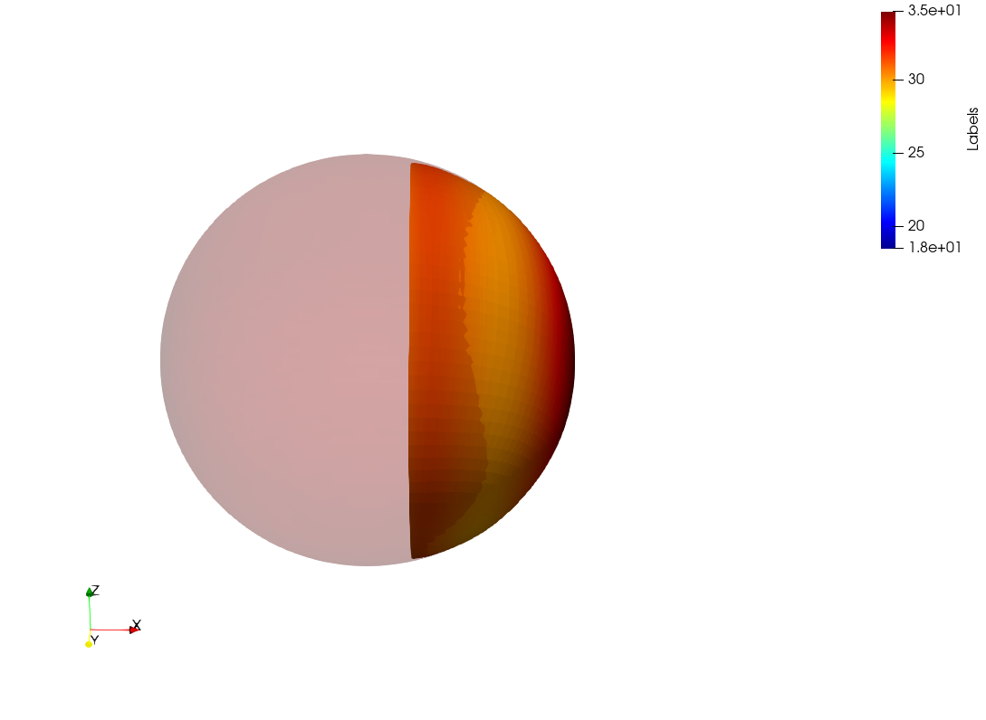}}
\hspace{0.05\textwidth}
\centering
\vspace{0.05\textwidth}
\centering
   \subfloat[Partial Varifold \\Local Regularization \label{fig:pv_regul_local}]{%
      \includegraphics[clip, width=0.3\textwidth, trim=4cm 3cm 4cm 3cm]{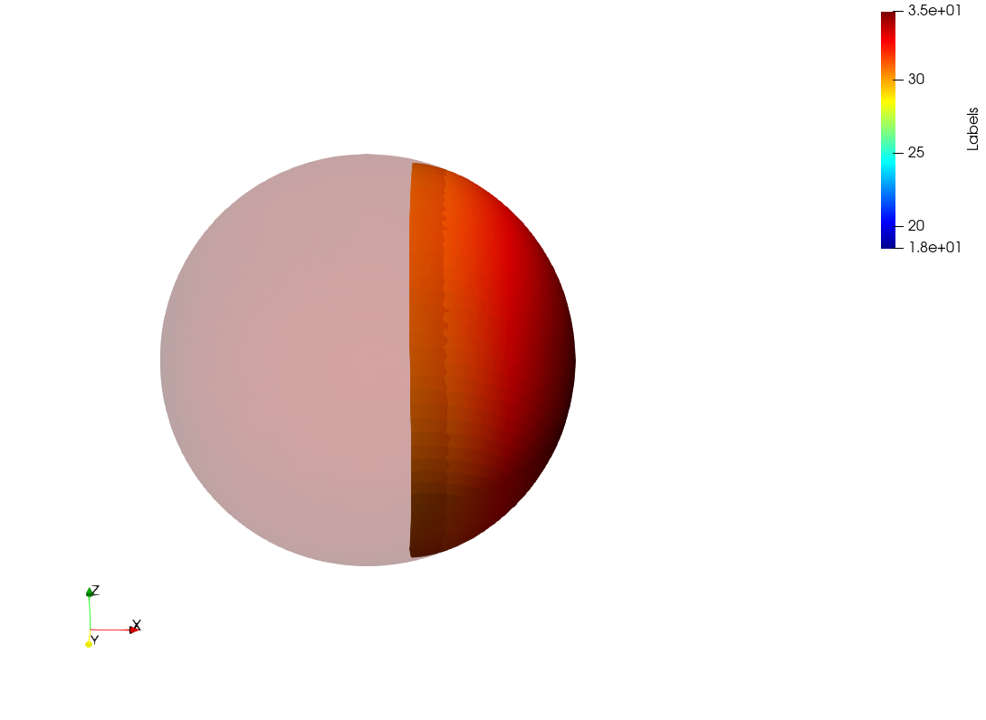}}
\end{minipage}\par\medskip
\caption{\label{fig:regularization} Influence of the regularization term on the non rigid deformation of a truncated sphere onto a complete one.
(a): Source (blue, opaque) and target (red, transparent) surfaces. 
(b-c-d): registration results.
The colormap for (b-c-d) indicates the euclidean distance (in mm) of the points to their initial position before diffeomorphic deformations. }
\end{figure}

This enforces the conservation of the norm of the deformed shape. Yet in practice it can lead to local deformations of one part of the shape compensated with another part (Fig.~\ref{fig:regularization}). We therefore introduce a \textit{local} regularization allowing to locally preserve the mass by enforcing the terms inside the integral to be close to $0$ everywhere:

\begin{eqnarray}\label{eq:local_regul}
R_{local}(\Src,\Phi(\Src))
=  \int_\Src \left(\om_\Src(\x) - \om_{\Phi(\Src)}\left(\Phi(\x)\right) \Det{\restr{d_x\phi}{\tau_x S}} \right)^2 dx \quad \mbox{ (\textbf{Local})}
\end{eqnarray}

\paragraph{Discrete formulation} Similarly to the partial matching term, we can write the regularization terms (both global and local) in the discrete setting:

\begin{eqnarray}
R_{global}(\Src,\Phi(\Src))
=  \left(\sum\limits_{i=1}^{N_\Src} \om_\Src(\x_i) - \sum\limits_{i=1}^{N_\Src} \om_{\Phi(\Src)}(\Phi(\x_i))  \right)^2  \quad \mbox{ (\textbf{Discrete Global})}
\end{eqnarray}

\begin{eqnarray}
R_{local}(\Src,\Phi(\Src))
=  \sum\limits_{i=1}^{N_\Src} \left(\om_\Src(\x_i) - \om_{\Phi(\Src)}\left(\Phi(\x_i)\right) \frac{\norm{\eta_{\Phi(\x_i)}\Phi(\Src)}}{\norm{\eta_{\x_i}\Src}} \right)^2  \quad \mbox{ (\textbf{Discrete Local})}
\end{eqnarray}

The overall function to minimize in this LDDMM setting is given by the formula: 

\begin{eqnarray}
J_{reg}(v) = \lambda_1 \int_0^1 \Vert v_t \Vert_V^2dt + \underline{\Delta}({\phi_1^v(\Src)},\Tgt) + \lambda_2.R(\Src, \phi_1^v(\Src))\,
\end{eqnarray}\label{eq:clinical_func}

with $R = R_{global}$ or $R_{local}$.

\begin{Proposition}
Let $\lambda_1>0$ and $\lambda_2>0$ be two fixed parameters. The regularized partial matching problem, which consists in minimizing over $L_V^2$ the function $J_{reg}$ (defined in Eq.\ref{eq:clinical_func}) has a solution.
\end{Proposition}

Similarly to \cite{Antonsanti2021}, the proof boils down to showing that the mapping 
$ v\mapsto A(v) = \underline{\Delta}({\phi_1^v(S)},T) + R_{local}(\Src,{\phi_1^v(\Src)})$ is weakly continuous on $L_V^2$. We use the same notations, and define $(v_n)$ a sequence in $L^2_V$, weakly converging to some $v\in L^2_V$.
We only need to show that $R_{local}(\Src,{\phi_1^{v_n}(\Src)}) \longrightarrow R_{local}(\Src,{\phi_1^v(\Src)})$.\\
To simplify we denote $ \Src_n=\phi_1^{v_n}(\Src)$, $\Src_*=\phi_1^{v}(\Src)$ and for any $\x \in \RxSd$, $\dps f_n(\x) = \omega_\Srcn(\x) - \int_T {\min}_\epsilon \left( 1, \dfrac{\om_{\Srcn}(\x)}{\om_\Tgt(\y)} \right) k(\x,\y) dy$ and $f_*(\x)$ likewise for $\Src_*$.

Using \cite{Antonsanti2021}, we have that $d_x\phi^n$ converge to $d_x\phi$, uniformly on $x\in \Src$ (\cite{Glaunes2005}). In addition $\om_{\Src_n}$ converges uniformly to $\om_{\Src_*}$.
We can deduce that $R_{local}(\Src,{\phi_1^{v_n}(\Src)}) - R_{local}(\Src,{\phi_1^v(\Src)}) \longrightarrow 0$.$\quad \qed$

\subsubsection{Influence of the Regularization Term}

We illustrate the influence of the regularization terms with the example of the registration of a truncated surface onto a complete one. To do so we perform a LDDMM registration using a small regularization parameter $\lambda_1 = 1000$ in the functional $J_{reg}$ and we set $\lambda_2=1$. The data attachment term we use is the one proposed for the partial matching in Section.~\ref{eq:discrete_pm}.
We observe that without any regularization, the non-rigid deformations lead to global shrinkage of the source shape. On the contrary the proposed regularizations both global and local prevent from such shrinkage. The global one though preserves the norm of the shape at the end of the diffeomorphic deformation and does not prevent from inconsistent local deformations. The most regular deformation is thus induced by the local regularization, which allows to locally control the value of the varifold associated to the deformed source. It should be noted that we work at a small scale of attachment to the data, and that this influences the regularization. 
In the rest of the paper, we will use this local mass preserving term for the clinical application (section \ref{sec:clinical}).

\subsubsection{Implementation Details}\label{sec:experiments}
In all the following experiments, we initialize the registration by aligning objects barycenters since no prior positioning is known in our applications.
To model non-rigid deformations, we define the reproducing kernel $K_V$ of $V$ to be a sum of Gaussian kernels: 
$$K_V(x,y) = \sum_s \exp\left(-\Vert x-y\Vert^{2}\;/\;(\sigma_0/s)^{2}\right)$$ 
where $s \in [1,4,8,16]$ and $\sigma_0$ is about half the size of the shapes bounding boxes.
For each set of experiments we use the same hyperparameters ($\sigma_0, \sigma_W, \lambda_1, \lambda_2$) to compare the influence of the regularization and for the clinical application. The parameter $\sigma_0$ controls the range of the non-rigid deformations produced by LDDMM, while $\sigma_W$ is associated to the spatial kernel used for the data attachment term in the space of varifolds, and denotes the reach of each varifold.
Our Python implementation makes use of the libraries PyTorch \citep{pytorch} and \href{https://www.kernel-operations.io/keops/index.html}{KeOps} \citep{JMLR:v22:20-275}, to benefit from automatic differentiation and GPU acceleration of kernel convolutions.

\subsubsection{Computational Cost}

To compute the registrations, we use a TITAN RTX Graphic Processing Unit. One rigid deformation of a $444\times512\times512$ voxel grid is computed in $9.10\times10^{-1}s$ when the diffeomorphic deformation takes $57.0$ seconds.
In terms of optimization, the ICP (on CPU) takes $4.64\times10^{-2}s$ for a source and a target of approximately $10^{4}$ points. On the same data the rigid registration guided with our partial matching dissimilarity function takes $9.63\times10^{-1}s$ and the LDDMM optimization takes $274s$.
This huge difference between LDDMM and the others methods comes from the number of parameters to optimize in its framework : the dimension of the space times the number of points in the source shape.

There are many possible ways to accelerate the LDDMM, from reducing the number of points in the source to the approximation of the diffeomorphisms with other deformations models. 
No matter the deformation, the computational cost of the partial matching dissimilarity function for the same data is $7.24\times10^{-3}s$.


\section{Clinical application : Feature-based Multi-modality Liver Volume Registration}
\label{sec:clinical}

Medical images are often acquired through different modalities, including ultrasound, computed tomography, and magnetic resonance imaging, each providing different and complementary information.
In this context, image registration allows physicians to obtain combined inputs from different imaging modalities using for instance image comparison or fusion. The latter has been shown to be valuable in image-guided procedures, yielding less complications and decreasing radiation dose \citep{Rajagopal2016}. This section is the clinical application of the work introducing the partial matching in the space of varifolds \citep{Antonsanti2021} and the extensions proposed in \ref{sec:partial_matching}.

\subsection{CT/CBCT Volume Registration}

Transcatheter directed liver therapies are part of the therapeutic arsenal of primary and secondary liver malignancies. 
The objective of these procedures is to locally treat the tumor and be as selective as possible (meaning placing the microcatheter used to inject the treatment as close as possible to the tumor) to preserve surrounding healthy tissues all while ensuring the destruction of the malignant cells. 

These minimally invasive procedures are performed by navigating through the patient's arteries under real time 2D angiography, acquired through an imaging device called C-arm. 
Additionally, the latter can perform a 200 degrees rotation to allow 3D reconstruction of the patient's anatomy, called Cone Beam Computed Tomography (CBCT) \citep{Tacher2015}, to obtain a ``live" 3D imaging of the patient at point of care. Performing CBCT during such procedure improves tumor detection and navigation guidance (Fig.~\ref{fig:CBCT}).

Classically, preprocedural diagnostic CT scan or MRI are reviewed by the interventional radiologist to plan the procedure accordingly. 
The preprocedural acquisitions provide information on the entire liver anatomy, tumor burden and tumor feeding arteries that are decisive for procedural planning, such as number of tumors to be treated in one session and the dose of therapeutic agent to inject.  
Contrary to CT or MRI, CBCT is performed during the procedure, and the operator can compare procedural CBCT to preprocedural CT (Fig.~\ref{fig:CT}) to ensure adequate treatment delivery.

While CBCT offers a superior spatial resolution compared to conventional CT scan, with intra-arterial injection of contrast agent providing a detailed visualization of the arteries, low contrast visibility is better in CT (Fig.~\ref{fig:CBCT}). 
CBCT can also be subject to several artifacts such as beam hardening and motion artifacts that might decrease the CBCT performance to visualize the tumor, which is key to selective and successful treatment. 
A major difficulty in the fusion of a CT volume with a CBCT one comes from the fact that, unlike CT, liver is only partially visible in CBCT (due to the limited size of the field of view in the latter modality). 
In addition the acquisitions are taken at different times, potentially several weeks apart, and with different patient stances introducing deformations of the liver.
For all these reasons, the two types of volumes are very different one from another as illustrated in Fig.~\ref{fig:illustration_modalities}.

We propose a registration method based on liver surfaces (one of the feature that is visible in both the CT and CBCT whatever the clinical acquisition protocol) providing a deformation of the entire volume.
To that extent we apply our partial matching dissimilarity term allowing to tackle the issue of partial correspondence between the truncated surface extracted from the CBCT (Live) and the one extracted from the CT (Liver).
The registrations are evaluated on landmarks inside the liver, which were annotated by a physician.

\subsection{Database Description}

The database is composed of CBCT/CT pairs where CBCT have been acquired during hepatic arteriography and CT scans obtained at early or late arterial phase.
We do not provide the acquisition parameters here, yet the spatial resolution (in $mm$) of the volumes are $(0.45,0.45,0.45)$ for the CBCT and in average $(0.75,0.75,1.25)$ for the CT.
Both were selected to show good visualization of the vessels and tumors.  
In total,  19 pairs of CT/CBCT liver volumes were evaluated, as the one illustrated in Fig.~\ref{fig:illustration_modalities}. 
\begin{figure*}[!t]
\centering
   \subfloat[\label{fig:POI_CBCT}]{%
      \includegraphics[clip, width=0.4\textwidth]{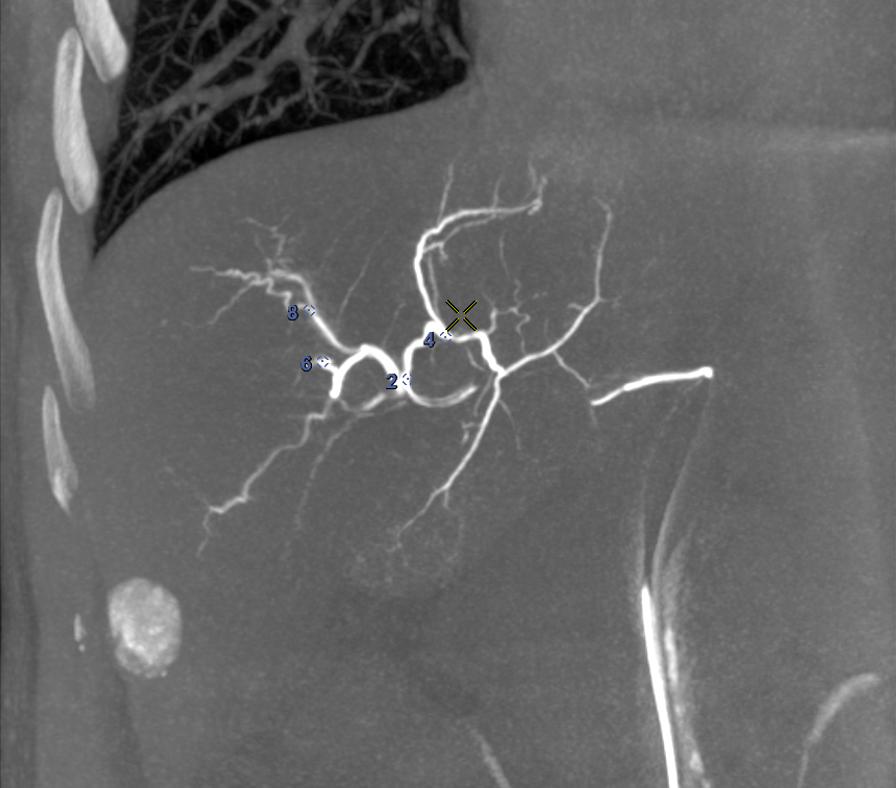}}
\hspace{.5cm}
   \subfloat[\label{fig:POI_CT}]{%
      \includegraphics[clip, width=0.4\textwidth]{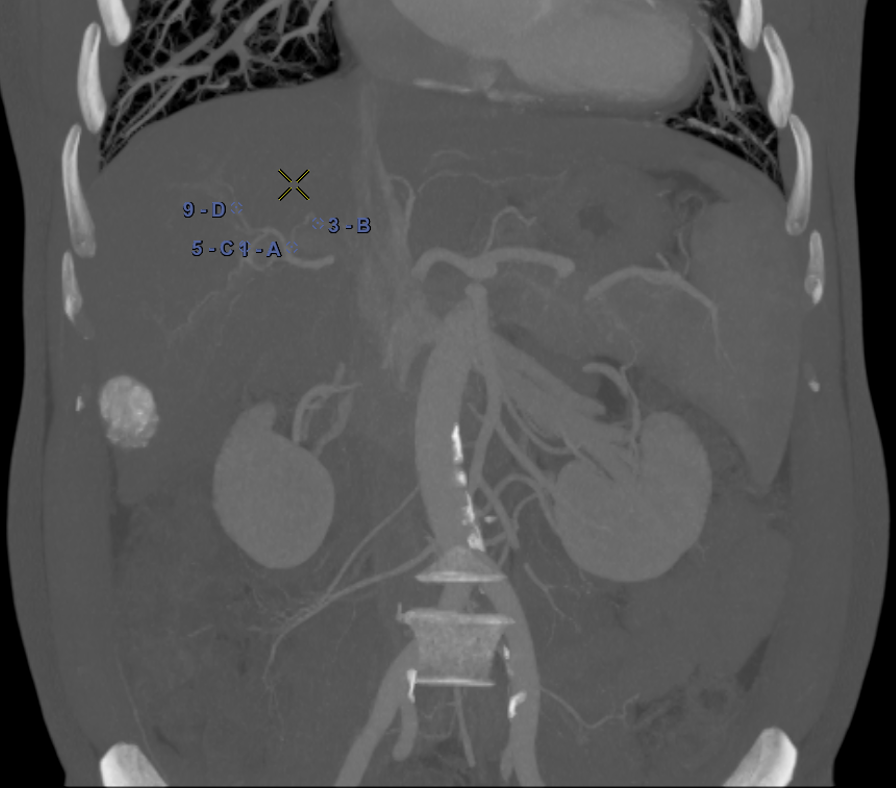}}
\hspace{.5cm}
   \subfloat[\label{fig:Lesion_CBCT}]{%
      \includegraphics[clip, width=0.4\textwidth]{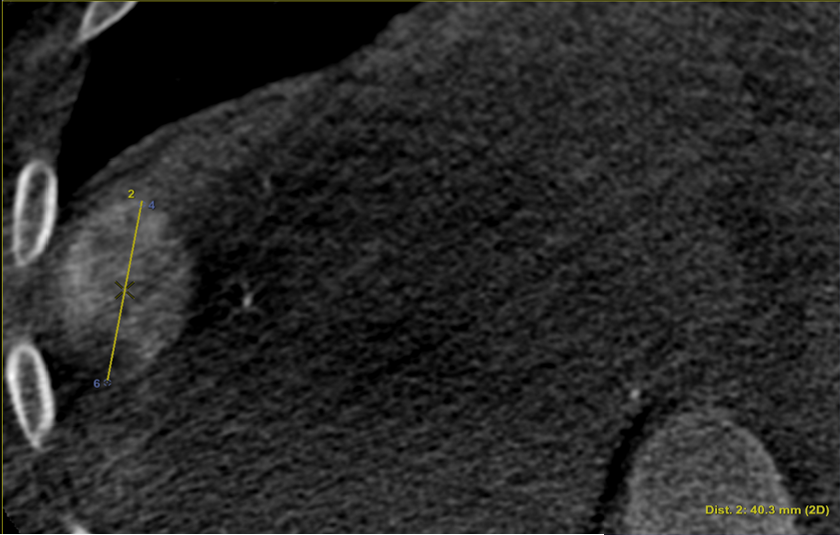}}
\hspace{.5cm}
   \subfloat[\label{fig:Lesion_CT}]{%
      \includegraphics[clip, width=0.4\textwidth]{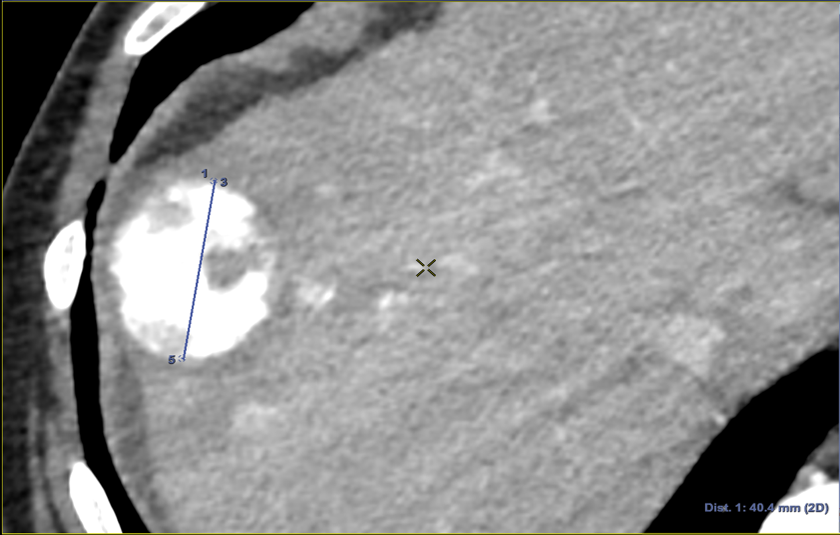}}

\caption{\label{fig:illustration_annotations} Examples of annotated Points of Interest on CBCT (a) and on CT (b). Examples of annotated longest axis diameter lesion visualized on CBCT (c) and on CT (d). All images come from the same patient.}
\end{figure*}

\subsubsection{Liver Segmentation} The livers were segmented in each modality using a deep neural network \citep{Milletari2016} providing a binary volume in both modalities.
In this application, the segmentations were evaluated by a clinical specialist and manually corrected if a major error was detected such as missing part of the liver. The idea is to be close to the clinical set up.
By doing so, we ensure that the registrations are based on features as reliable as possible.
The mesh of the surfaces were then extracted and decimated, leading to meshes of approximately $10^4$ points per surface. In the case of CBCT volumes, the meshes are then cut by the cylinder corresponding to the field of view using \cite{Musy2021_vedo}. The CBCT livers surfaces obtained are then truncated.

\subsubsection{Points of Interest} For each patient, the branches of the proper hepatic artery visible on CT volume were annotated with Points of Interest (POIs) for evaluation purpose. Each POI was similarly annotated in the same location on the corresponding CBCT volume. Selected POIs often corresponded to arterial bifurcations that are easily identifiable on both CT and CBCT. For each pair of volumes, a physician annotated 10 POIs (Fig.~\ref{fig:POI_CBCT},\ref{fig:POI_CT}). 
Because of the limited visibility of distal hepatic arteries on arterial phase CT compared to CBCT acquired during hepatic arteriography, most of POIs were located close to the bifurcation of the proper hepatic artery, thus mainly located at central parts of the livers, of importance to physicians.

\subsubsection{Tumors Annotation} To evaluate the registrations, in addition to POIs, we annotated the longest axis diameter of a tumor according to \cite{Ghosn2020} to ensure better reproducibility of the annotations across volumes (Fig.~\ref{fig:Lesion_CBCT}, Fig~\ref{fig:Lesion_CT}). 
It was done in the axial view of the volumes for tumors that were visible in both modalities. 
The axis can be decomposed into 3 Tumor Points : the extremities and the center. 
In the database, one invasive tumor could not be annotated, reducing the number of pairs of tumors to 18.
The annotated tumors were located in all the liver segments and their size varied from 9mm to 109mm.
This variability in terms of position and size provides a complementary information to that of the POIs.

\subsection{Liver Surface Registration with Partial Matching}\label{subsec:surface_registration}

\begin{figure*}[!ht]
\centering
   \subfloat[\label{fig:source}]{%
      \includegraphics[trim={0cm 2cm 4cm 0cm}, clip, width=0.35\textwidth]{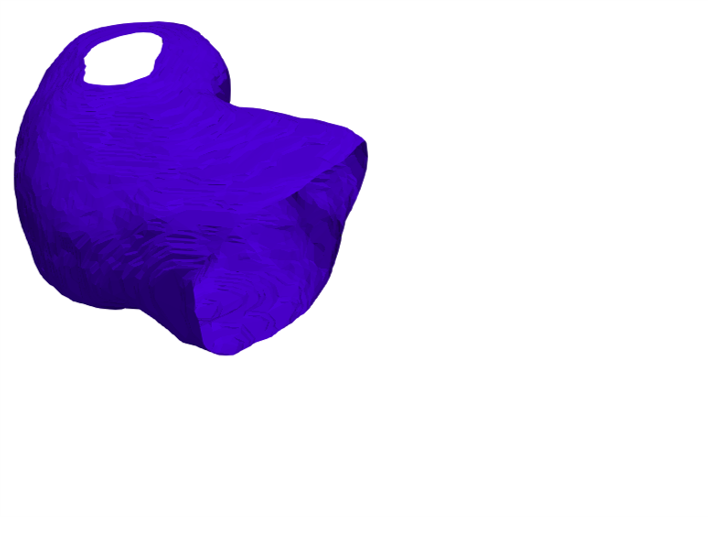}}
\hspace{.5cm}
   \subfloat[\label{fig:target}]{%
      \includegraphics[trim={0cm 2cm 4cm 0cm}, clip, width=0.35\textwidth]{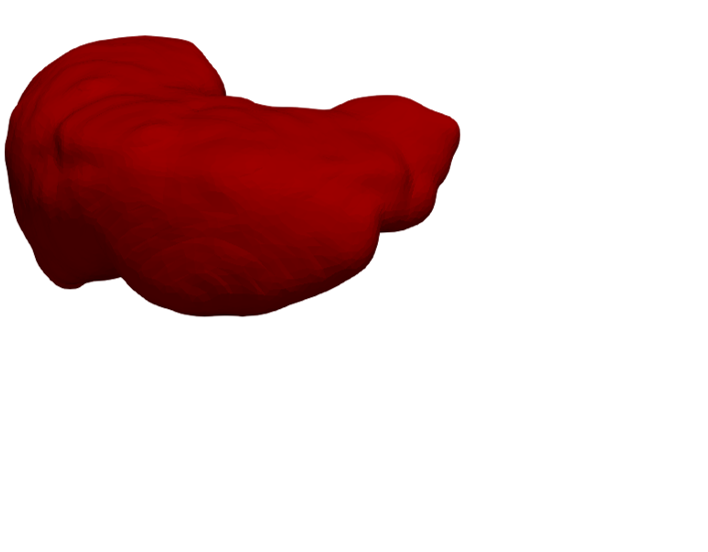}}
\par
\vspace{0.05\textwidth}
\centering
   \subfloat[\label{fig:jac_varif}]{%
      \includegraphics[trim={0cm 2cm 4cm 0cm}, clip, width=0.35\textwidth]{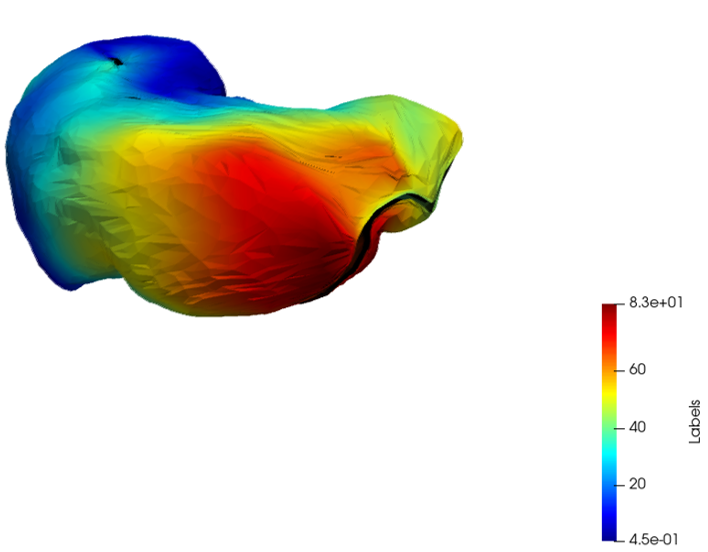}}
\hspace{.5cm}
\centering
   \subfloat[\label{fig:jac_partial_varif}]{%
      \includegraphics[trim={0cm 2cm 4cm 0cm}, clip, width=0.33\textwidth]{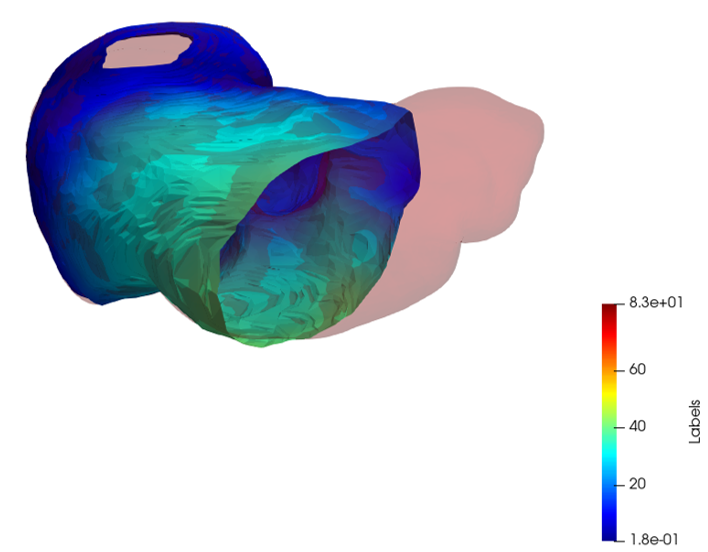}}
\hspace{.5cm}
\centering
   \subfloat[\empty \label{fig:colorbar}]{%
      \includegraphics[trim={14cm 0cm 0.5cm 7.5cm}, clip, width=0.2\textwidth]{./local_regul_local_PV.png}}\\

\caption{\label{fig:Surface_partial_registration_jacob} Registration of a truncated liver's surface (Live) from a CBCT (a) onto a complete liver's surface (Liver) from a CT (b) for Patient 2. Varifold registration (c);  Partial normalized registration (Def.~\ref{def:partial_normalized_dissimilarity}) with a local regularization (Eq.~\ref{eq:local_regul}) (d). The color scale indicates the euclidean distance (in mm) of the points to their initial position before diffeomorphic deformation.}
\end{figure*}

As a first registration step, the truncated liver surfaces from  CBCT were registered onto complete liver surfaces from CT scans with a LDDMM deformation model using the discrete framework described in Sec.~\ref{sub:discrete}.
The LDDMM deformations of the truncated livers surfaces with partial dissimilarity function may lead to small shrinkage of the borders.
To compensate this phenomenon in the application we added the a priori regularization of Eq.~\ref{eq:local_regul} to the partial data attachment term that prevents from strong local deformations. 
For illustration purpose, one subject was registered twice with this model: once with the distance in the space of Varifolds, once with the normalized partial dissimilarity term (Def.~\ref{def:partial_normalized_dissimilarity}) with the local regularization (Eq.~\ref{eq:local_regul}). This particular experiment is illustrated in Fig.~\ref{fig:Surface_partial_registration_jacob}.
Both results are generated with the same deformations and regularization parameters. 
As expected the Varifold distance leads to unrealistic deformations that tend to fill the holes in the source shape to cover the entire target. 
From the anatomical and medical point of view this is misleading and can not be used in a clinical application of multi-modality volumes registration. 
On the contrary the partial matching produces a more realistic deformation of the source onto a subset of the complete surface. We will only use and discuss this model in the following.

\paragraph{} 
To initialize the LDDMM deformations, one classically performs a rigid registration. We find in the literature that the livers are principally deformed in translation, so we tested a set of combinations between rigid deformations and LDDMM. We selected the methods providing the best results: a translation followed by LDDMM (denoted \textit{translation+LDDMM}) and a rigid deformation of limited angulation followed by LDDMM (denoted \textit{rigid+LDDMM}).
The rigid registration is limited to rotations between $-15^{\circ}$ and $15^{\circ}$ around each axis that is the range of realistic rotations for the liver deformations. 

In addition to these registration methods, we tested the standard rigid Iterative Closest Point (ICP) applied to the surfaces, using as data attachment term the function:
\begin{eqnarray*}
\Delta_{ICP}(\Src,\Tgt)
= \dfrac{1}{Card(\Src)}\sum_{x \in \Src}\min_{y \in \Tgt}\left(\norm{x-y}_{\Rd} \right). 
\end{eqnarray*}\label{eq:semi_hsdf}
By minimizing this term one minimizes the average distance of the source points to the target. This asymmetric function can be seen as a partial matching dissimilarity term, being equal to $0$ if $\Src$ is included in $\Tgt$.

\subsubsection{Implementation details} 

LDDMM is computed with the partial matching dissimilarity term and the localized mass preservation term Eq.~\ref{eq:local_regul}. In the rest of the paper $J$ is written as follows:

\begin{eqnarray}
J(v) = \lambda_1 \int_0^1 \Vert v_t \Vert_V^2dt + \underline{\Delta}({\phi_1^v(\Src)},\Tgt) + \lambda_2 R_{local}(\Src, \phi_1^v(\Src))\,
\end{eqnarray}\label{eq:clinical_func_recall}

The optimization of this functional $J$ is performed using Limited-memory Broyden–\-Fletcher–\-Goldfarb–\-Shanno (L-BFGS) algorithm. 
In the LDDMM framework the cost of the deformations is controlled by the parameters $\lambda_1$ and $\lambda_2$. To enforce smooth deformations of the surface as well as its ambient space, we set $\lambda_1$ to $10^7$, and we control the risk of shrinkage of the partial matching non-rigid registration by setting $\lambda_2 = 1$. 
The Reproducing Kernel of the deformations is the same as in~\ref{sec:experiments} allowing large deformations of the shapes along with more detailed ones. 
To better register the shapes, we also use a multi-scale registration scheme for the data attachment terms by iterative applications with $\sigma_W=10mm$ and $\sigma_W=5mm$ the output of the optimization at scale $10$ is used as input at the scale $5$.

\subsection{From Surface to Volume Registration}\label{subsec:volume_registration}

Both rigid deformation and LDDMM can be extended to the whole volume. 
\begin{figure*}[!t]
\begin{minipage}{.99\linewidth}
\centering
\begin{minipage}{.31\linewidth}\centering \small Rigid \end{minipage}
\hfill\begin{minipage}{.32\linewidth}\centering \small Rigid+LDDMM \end{minipage}
\hfill\begin{minipage}{.32\linewidth}\centering \small Translation+LDDMM  \end{minipage}
\begin{minipage}{.31\linewidth}\centering \small with ICP \end{minipage}
\hfill\begin{minipage}{.32\linewidth}\centering \small with Partial Varifold \end{minipage}
\hfill\begin{minipage}{.32\linewidth}\centering \small with Partial Varifold \end{minipage}
\vspace{-5mm}
\centering
   \subfloat[POIs metric: 2.77mm, \\ Tumor metric : 9.12mm\label{fig:sagi_icp_56}]{%
      \includegraphics[clip, width=0.29\textwidth]{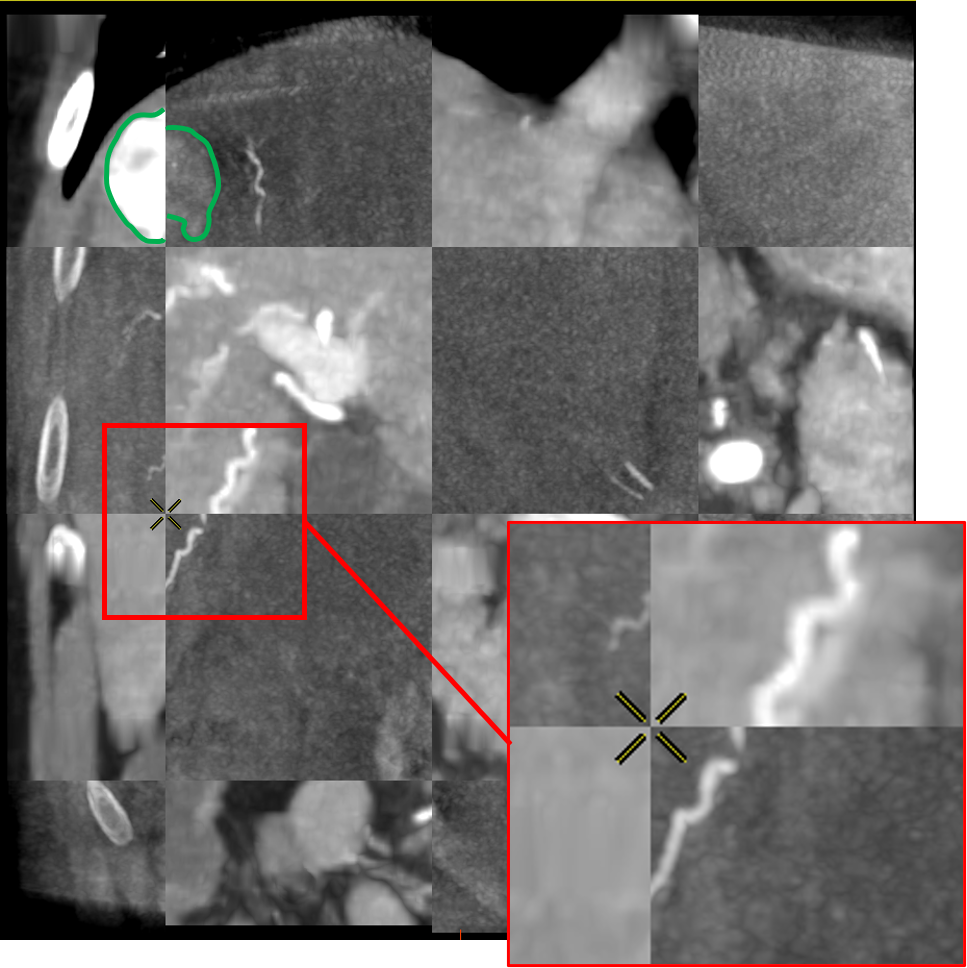}}
\hspace{0.05\textwidth}
\centering
\centering
   \subfloat[POIs metric: 3.61mm, \\ Tumor metric : 9.32mm\label{fig:sagi_atap_56}]{%
      \includegraphics[clip, width=0.29\textwidth]{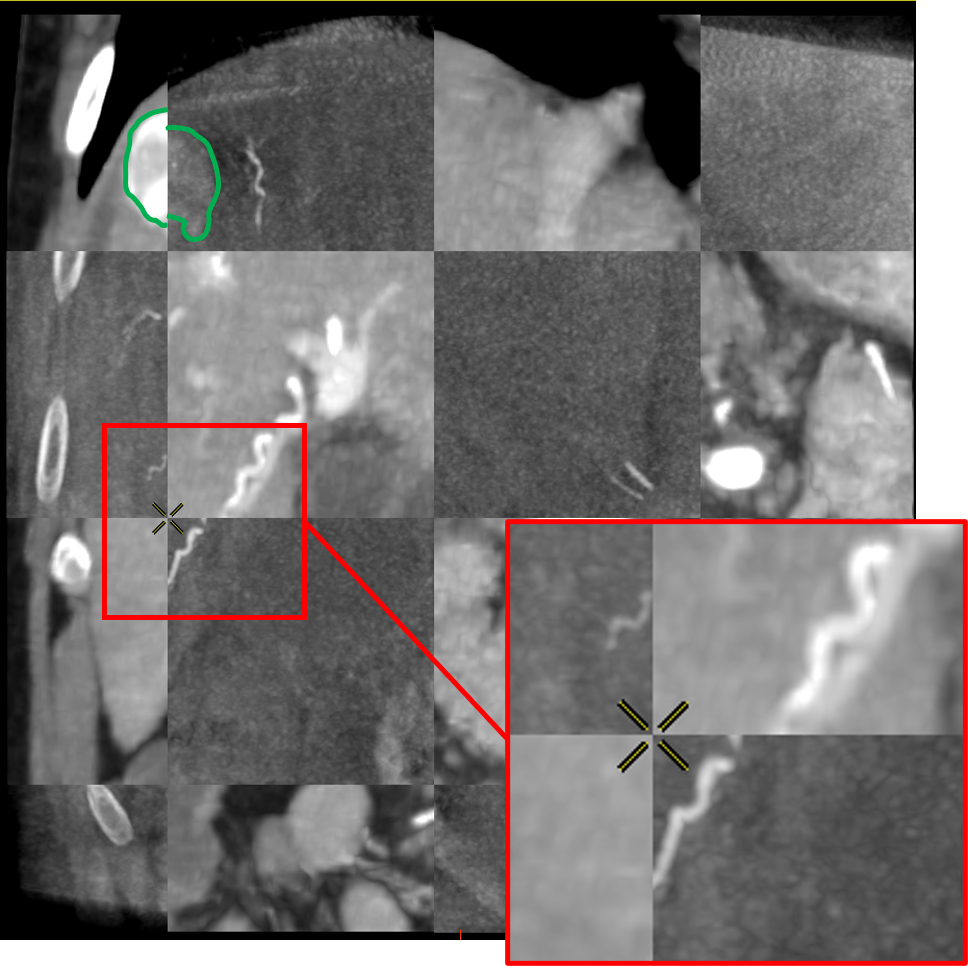}}
\hspace{0.05\textwidth}
\centering
\centering
   \subfloat[POIs metric: 4.22mm, \\ Tumor metric : 2.86mm\label{fig:sagi_arap_56}]{%
      \includegraphics[clip, width=0.29\textwidth]{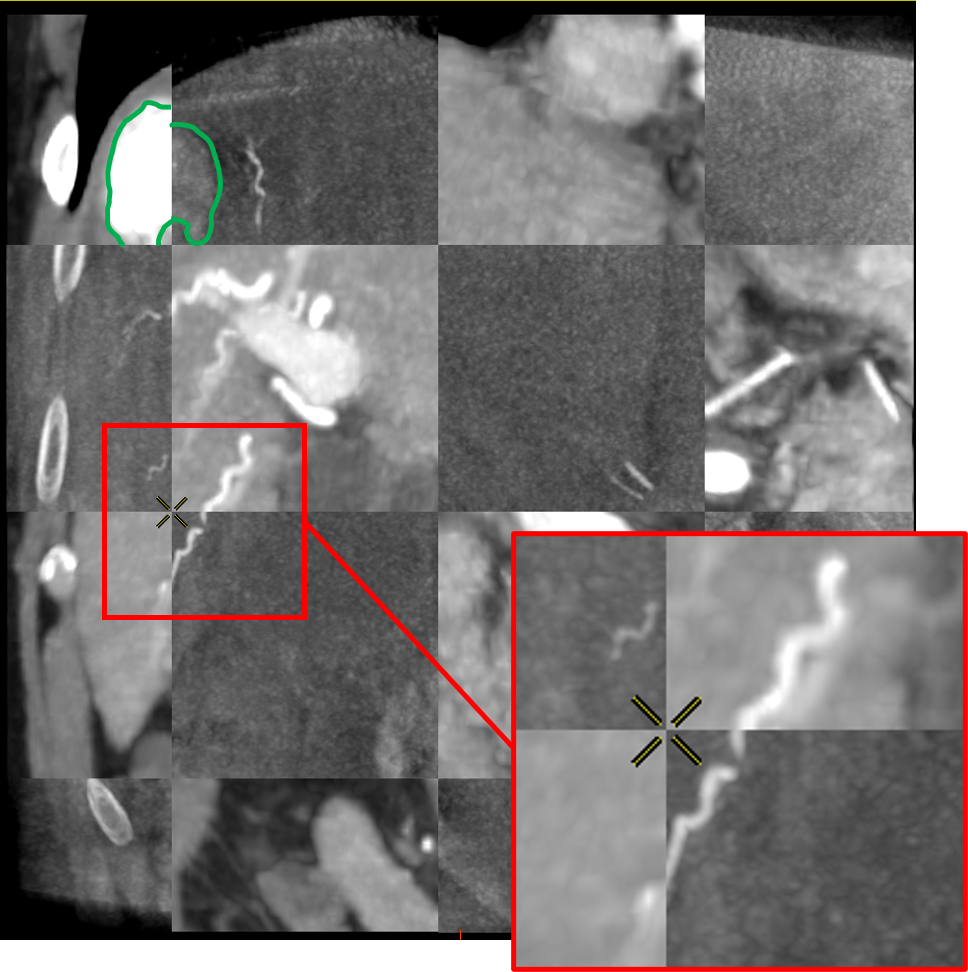}}
\end{minipage}\par\medskip
\caption{\label{fig:volume_comparison_56} Tiled visualization of the registrations for Patient 11 for which the approaches using rigid deformations (a,b) register correctly the vessels but fail to align the tumor. The tiles of the CBCT target volume are the dark ones, those of the deformed CT volumes are the light ones.}
\end{figure*}
To do so, the voxel grid of the source volume is deformed and interpolated on the target volume. 
In the case of LDDMM deformations, the voxel grid is deformed with the diffeomorphism of $\mathbb{R}^3$ as described in~\ref{sec:LDDMM}.
The values of the interpolated grid are then reported in the initial volume, providing a registration of the CT volume onto the CBCT one as illustrated in Fig.~\ref{fig:volume_comparison_56}. 

\begin{figure*}[!t]
\begin{minipage}{.99\linewidth}
\centering
\centering
\begin{minipage}{.31\linewidth}\centering \small Rigid \end{minipage}
\hfill\begin{minipage}{.32\linewidth}\centering \small Rigid \end{minipage}
\hfill\begin{minipage}{.32\linewidth}\centering \small Translation+LDDMM  \end{minipage}
\begin{minipage}{.31\linewidth}\centering \small with ICP \end{minipage}
\hfill\begin{minipage}{.32\linewidth}\centering \small with Partial Varifold \end{minipage}
\hfill\begin{minipage}{.32\linewidth}\centering \small with Partial Varifold \end{minipage}
\vspace{-5mm}
\centering
   \subfloat[POIs metric: 9.75mm\label{fig:sagi_icp_04}]{%
      \includegraphics[clip, width=0.28\textwidth]{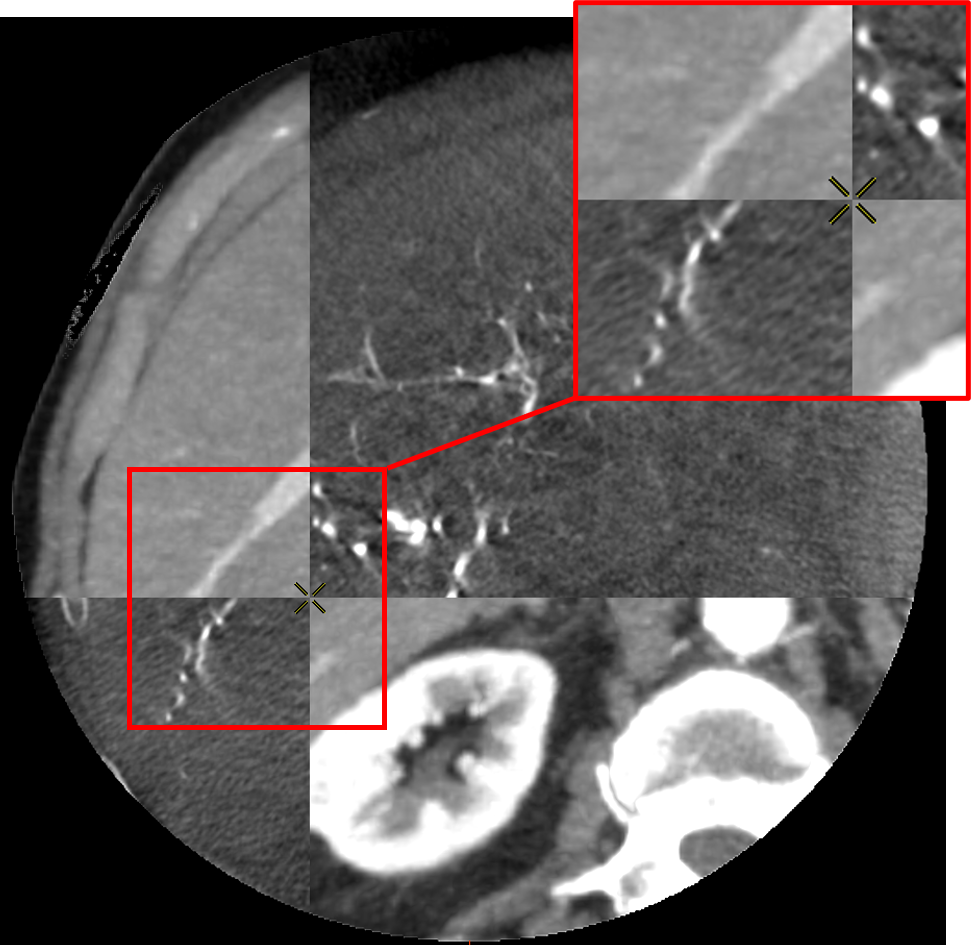}}
\hspace{0.05\textwidth}
\centering
\centering
   \subfloat[POIs metric : 8.53mm\label{fig:sagi_atap_04}]{%
      \includegraphics[clip, width=0.28\textwidth]{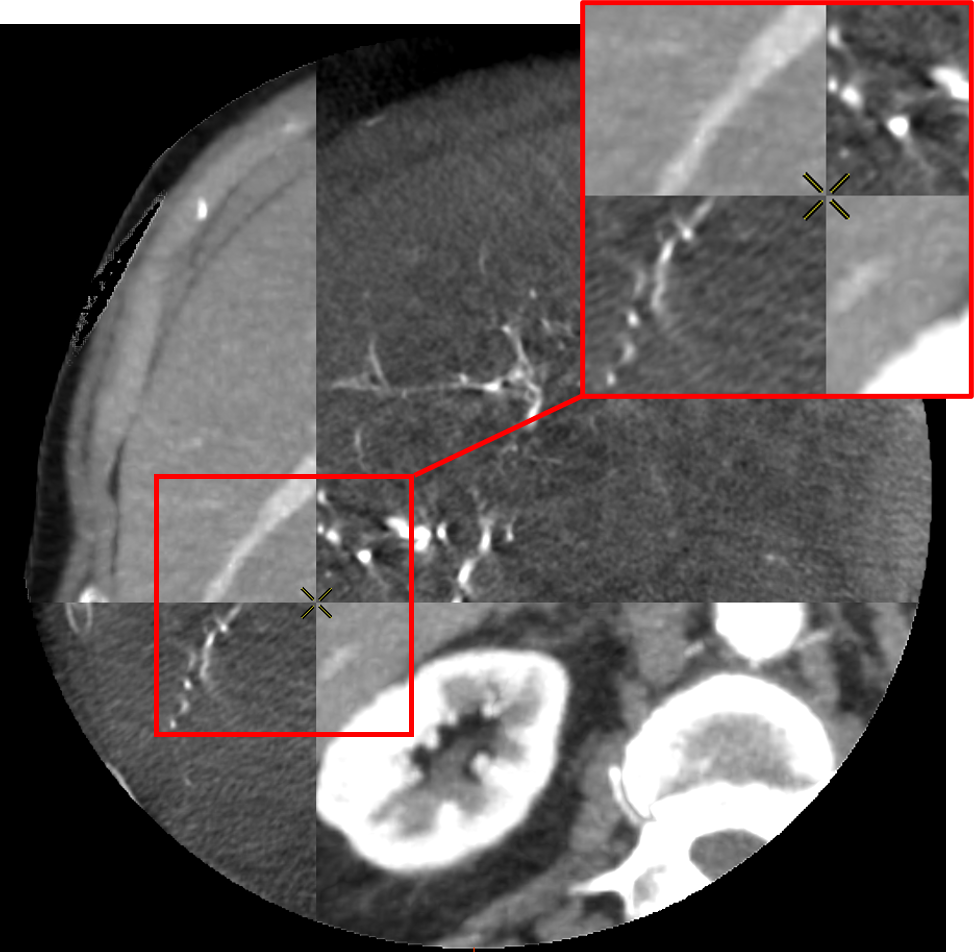}}
\hspace{0.05\textwidth}
\centering
\centering
   \subfloat[POIs metric : 4.94mm\label{fig:sagi_arap_04}]{%
      \includegraphics[clip, width=0.28\textwidth]{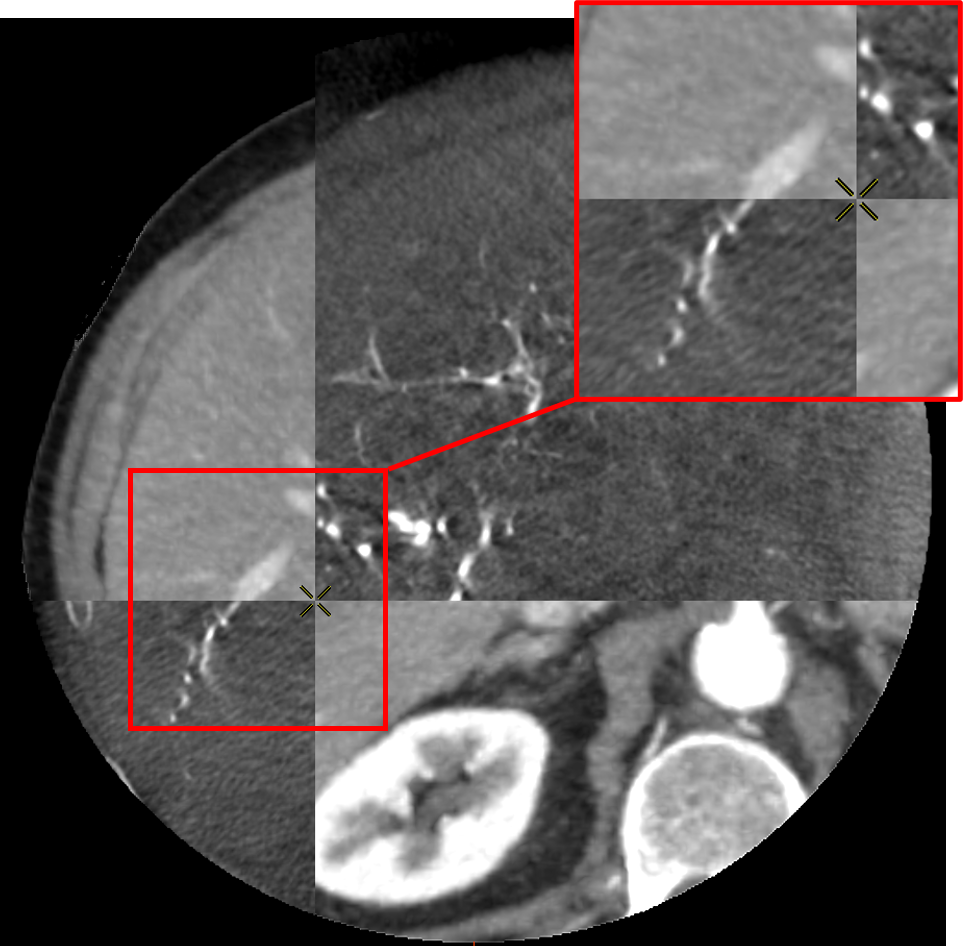}}
\par
\vspace{0.05\textwidth}
\centering
   \subfloat[Tumor metric : 19.8mm\label{fig:axial_icp_04}]{%
      \includegraphics[clip, width=0.28\textwidth]{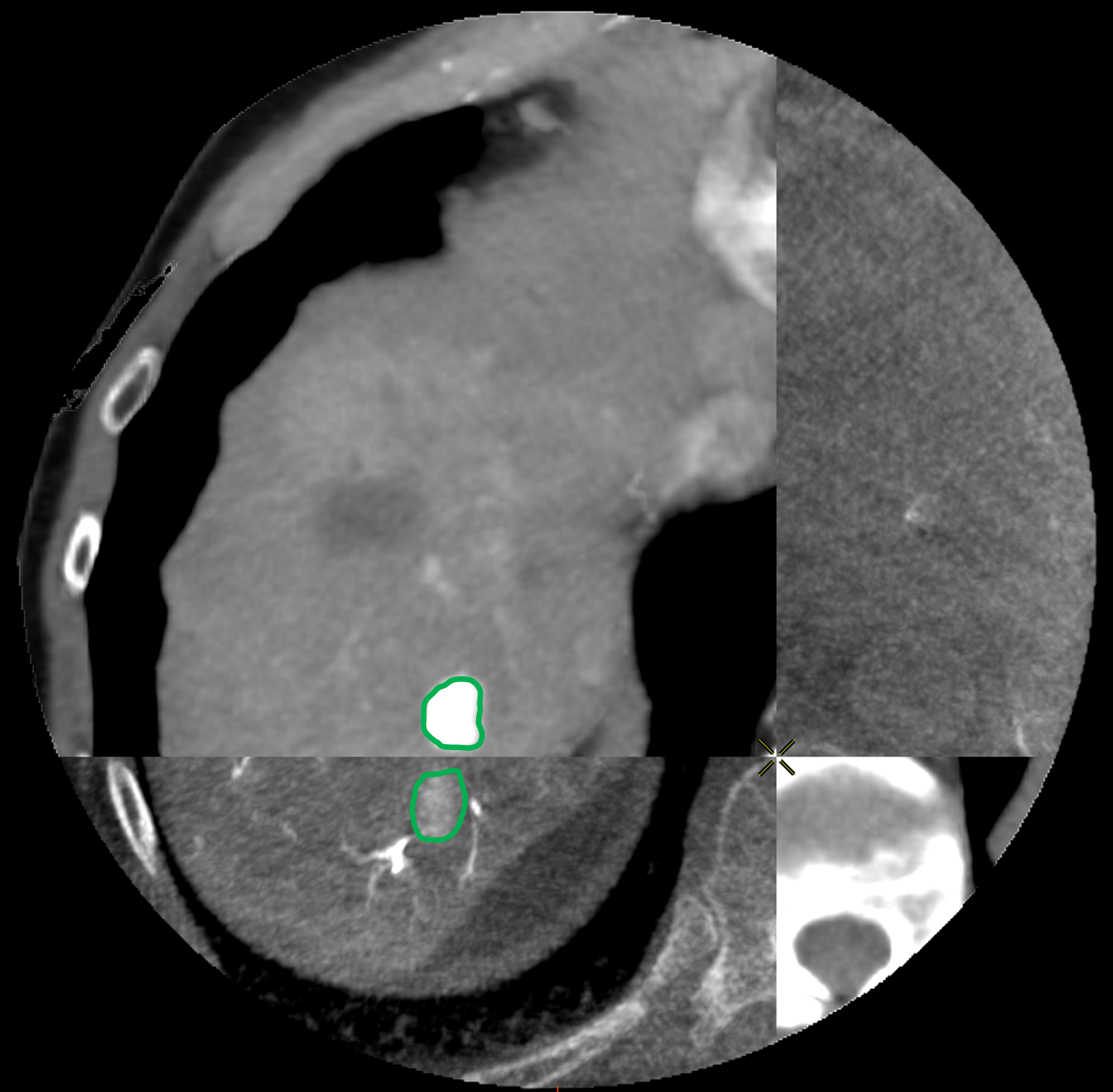}}
\hspace{0.05\textwidth}
\centering
\centering
   \subfloat[Tumor metric : 17.4mm\label{fig:axial_atap_04}]{%
      \includegraphics[clip, width=0.28\textwidth]{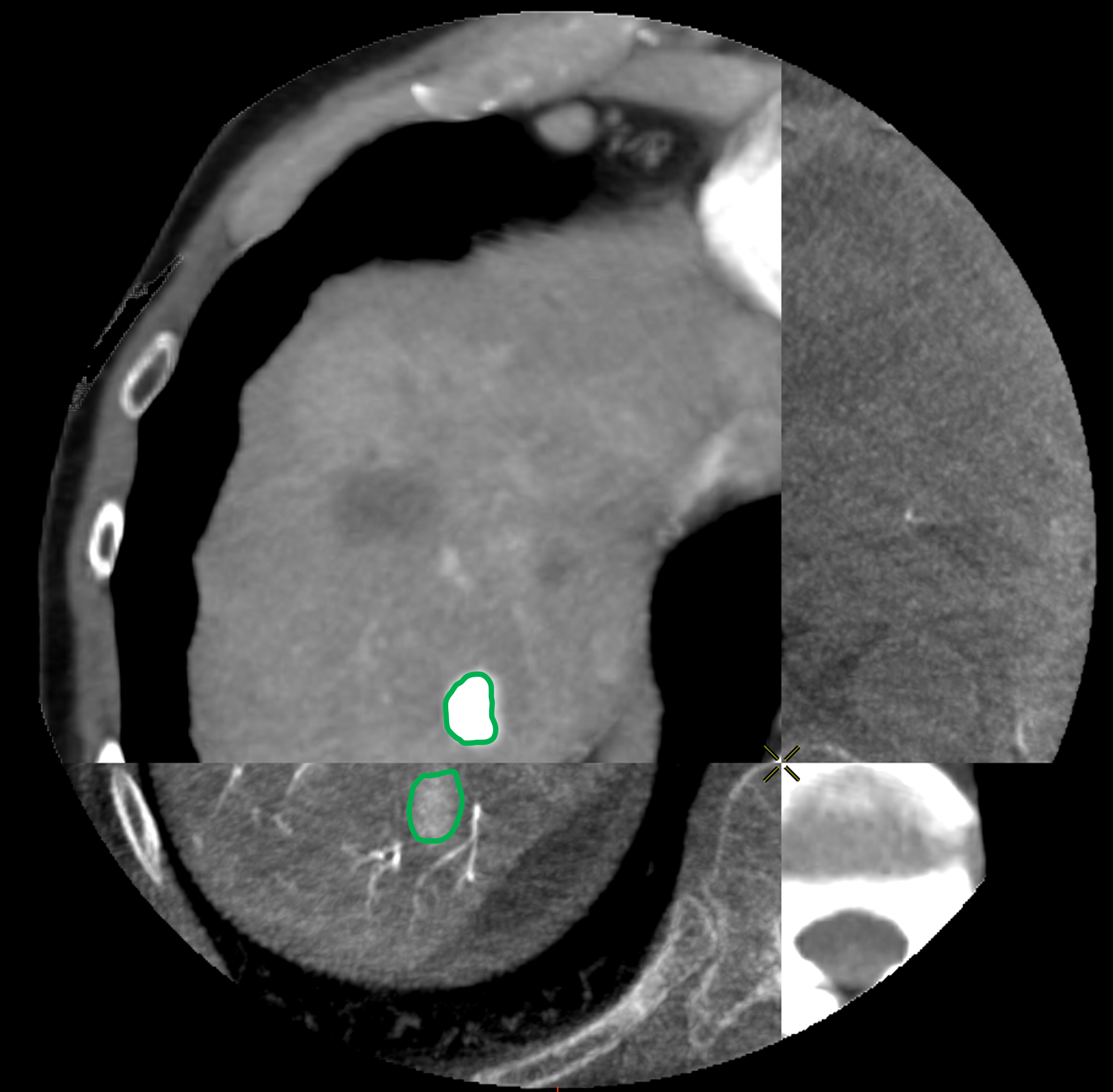}}
\hspace{0.05\textwidth}
\centering
\centering
   \subfloat[Tumor metric : 3.62mm\label{fig:axial_arap_04}]{%
      \includegraphics[clip, width=0.27\textwidth]{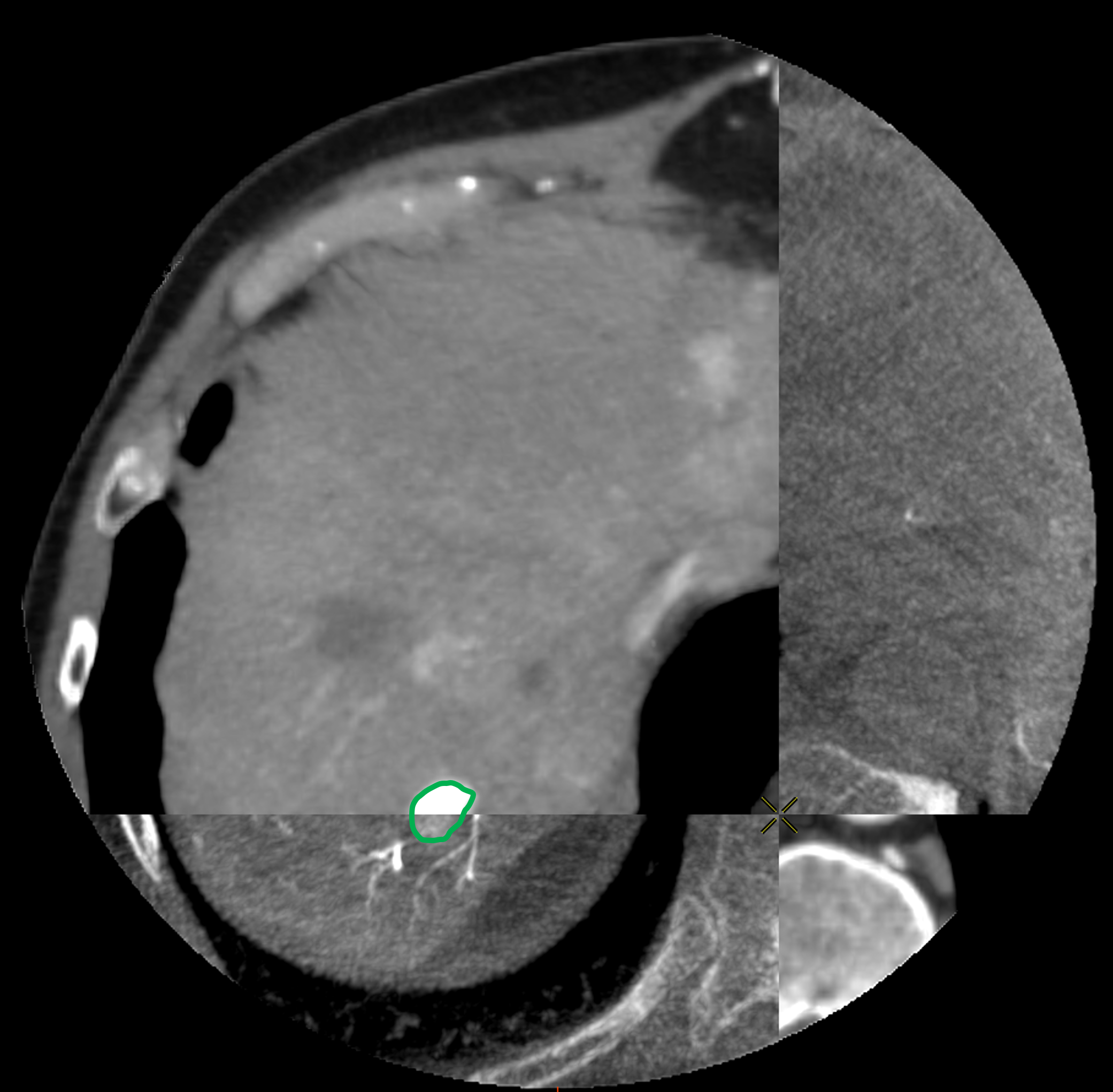}}
\end{minipage}\par\medskip
\caption{\label{fig:volume_comparison_04} Tiled visualization of the registrations for Patient 2. The tiles of the CBCT target volume are the dark ones, those of the deformed CT volumes are the light ones. First row : sagittal view, second row : axial view.}
\end{figure*}

The registrations of the volumes can be visualized to qualitatively assess the registration in the livers. 
To provide a 2D visualization of the results, we use in Fig.~\ref{fig:volume_comparison_56}  and Fig.~\ref{fig:volume_comparison_04} a \textit{tiled} representation that alternatively shows two volumes. 
Such visualization allows to see the continuities between the volumes. 
Each tile contains a 2D view of the CT volume (light tiles) or the CBCT one (dark tiles).
As the dynamics of the images are very different, and the tissues that emerge differ from one modality to another, we are interested in the continuity of the emerging structures such as vessels, liver parenchyma or tumors. 

\subsection{Evaluation and Results}

We recall that the key to clinical success is to register precisely the local area around the tumor despite the fact that this tumor is not segmented in CBCT (live) clinical routine (and thus not usable in the registration procedure).
Therefore the liver registration is evaluated through the euclidean distance between the deformed points of interest of the CBCT and those of the CT. It is done similarly with the tumors landmarks. Since the POIs are more centered than the tumor landmarks (see Fig.\ref{fig:illustration_annotations}), this second evaluation is significant as we will further explain in the discussion. 
The detailed results per case are provided in Table~\ref{tab:poi_details} and Table~\ref{tab:lesions_details}. In addition, we evaluate the registration at the surface level, which provides an indication of the overall registration quality, and gives the physician an additional benchmark for the comparison of CT and CBCT volumes. 
We first compute the barycenters registrations between the surfaces and apply the resulting translations to the volumes.We obtain an average distance of $18.3mm$ between the POIs, and of $21.5mm$ between the tumors landmarks. 
We show these values as red lines in the corresponding figures. They illustrate what the physicians can quickly obtain during the procedures, registering the volumes in translation by clicking one corresponding point in both modalities.
As a reference method for the rigid registration, we also computed the ICP registration directly based on the distance between the points of interest. 
This registration setting is the only one to exploit the annotated landmarks as input and it will only be used for quantitative comparison.

Therefore the liver registration is evaluated through the euclidean distance between the deformed points of interest of the CBCT and those of the CT. It is done similarly with the tumors landmarks. Since the POIs are more centered than the tumor landmarks (see Fig.\ref{fig:illustration_annotations}), this second evaluation is significant as we will further explain in the discussion. The detailed results per case are provided in Table~\ref{tab:poi_details} and Table~\ref{tab:lesions_details}. In addition, we evaluate the registration at the surface level, which provides an indication of the overall registration quality, and gives the physician an additional benchmark for the comparison of CT and CBCT volumes. We first compute the barycenters registrations between the surfaces and apply the resulting translations to the volumes.We obtain an average distance of $18.3mm$ between the POIs, and of $21.5mm$ between the tumors landmarks. We show these values as red lines in the corresponding figures. They illustrate what the physicians can quickly obtain during the procedures, registering the volumes in translation by clicking one corresponding point in both modalities.

\subsubsection{Evaluation on the POIs} 
\begin{figure*}[!t]
\centering
   \label{fig:s_poi}%
      \includegraphics[clip, width=0.9\textwidth]{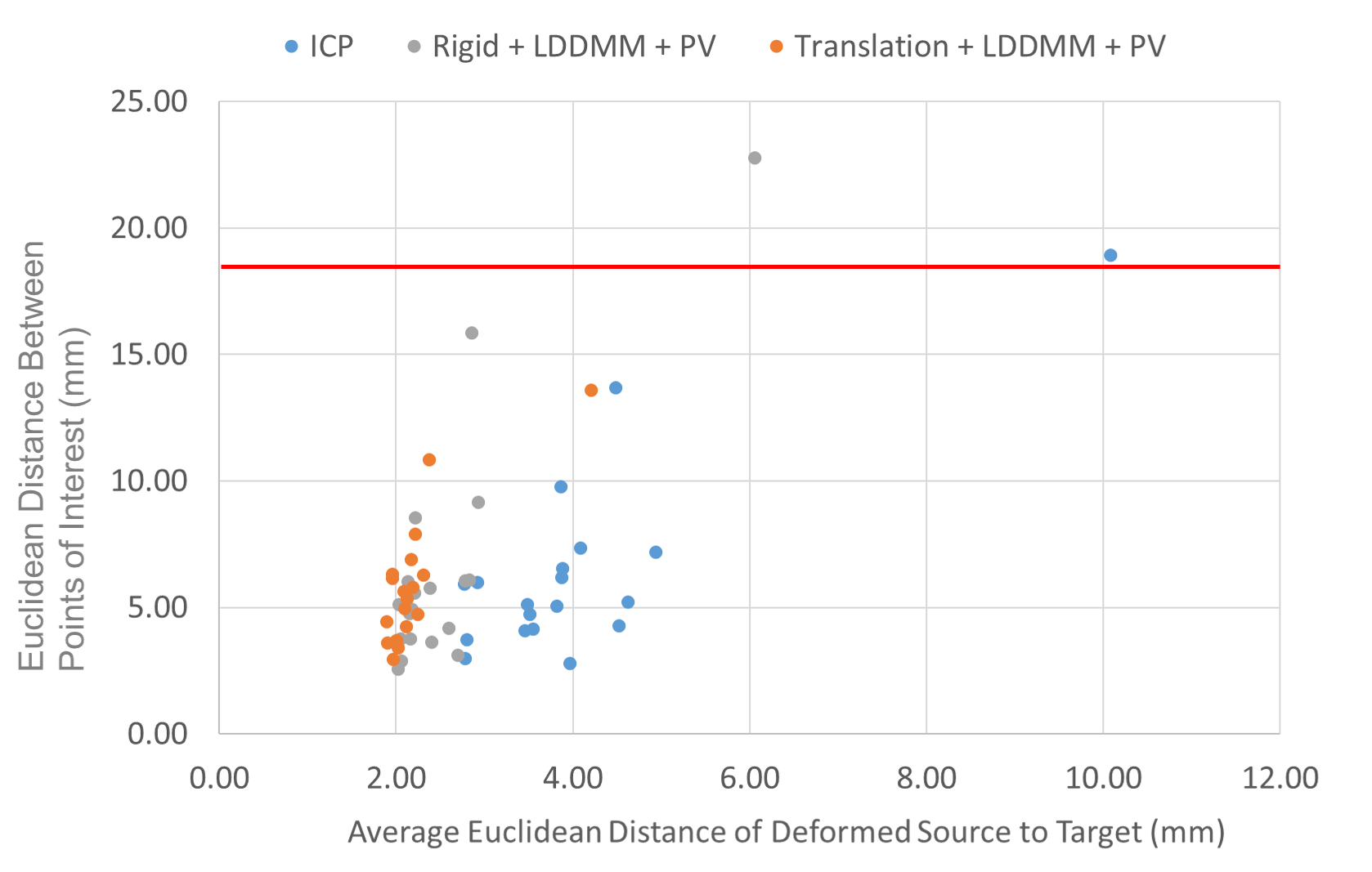}
\caption{\label{fig:scatter_poi} Registration's evaluation on the Points of Interest. The non-rigid LDDMM deformations based on partial matching allow robust surface registration while ensuring consistent deformation of the POIs. The red line corresponds to the average metric on POIs using barycenters registration.}
\end{figure*}

\begin{figure*}[!t]
\centering
   \label{fig:box_poi}%
      \includegraphics[clip, width=0.9\textwidth]{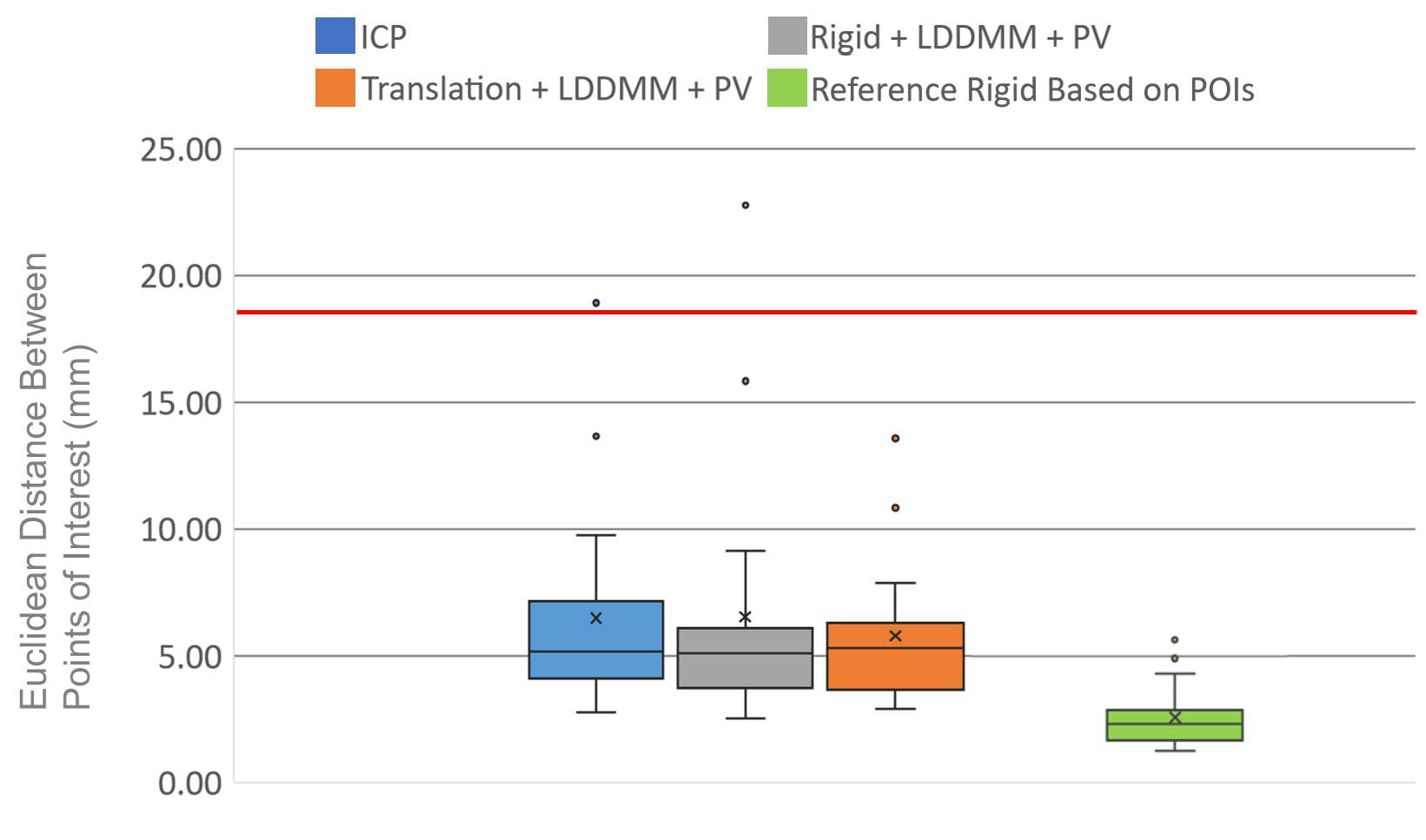}
\caption{\label{fig:results_poi} Registration's evaluation on the Points of Interest. The rightmost box corresponds to the reference rigid registration of the POIs, hence the best possible results for rigid deformations. The red line corresponds to the average metric on POIs using barycenters registration.}
\end{figure*}

We computed the euclidean distance between the POIs in the target volumes and the deformed ones. 
The results are presented in two figures: the first one (Fig.~\ref{fig:scatter_poi}) provides a detail of the distances between the POIs as a function of the average distances from the points of the deformed surface to those of the target surface. This gives an idea of the distance between the edges of the liver after registration and the influence on the distance between the POIs. The associated box plots in  Fig.~\ref{fig:results_poi} provide a summary of the interest point registration results according to the method used.
The right box in Fig.~\ref{fig:results_poi} corresponds to the ICP rigid registration based on the POIs for the data attachment term and is used as reference. Note that this box shows that the variations cannot be explained by rigid deformations only.
We observe in Fig.~\ref{fig:scatter_poi} that the LDDMM deformations allow a consistent and robust registration of the surfaces with an average distance of the deformed source points to the target of about $2mm$ in average. This cannot be achieved by only rigid deformations guided by ICP ($4mm$ in average), but must be validated with other metric to assess the quality of the deformation applied to the whole livers volumes. 
In terms of POIs distances, none of the three methods illustrated in this scatter plot shows significant difference with the others, as validated in Fig.~\ref{fig:results_poi}.

Each of them performs differently depending on which patient they are evaluated as one can see in the detailed table in Appendix~\ref{annex:poi_details}, but none of them stands out for the POIs metric. 
The best average performance $5.78mm\pm5.32$ is achieved with the translation+LDDMM deformation guided by our partial dissimilarity term yet it is not significantly better that the rigid ICP ($6.49mm\pm5.18$) or the rigid+LDDMM method ($6.54mm\pm5.09$).
When referring to the details in Appendix~\ref{annex:poi_details}, we see that the LDDMM deformations significantly improve the translations (reducing by $44\%$ the distance between the POIs). However, rigid registrations provide a poor initialization for the LDDMM deformations. 
By looking at the rotations angles obtained with the ICP based on the surfaces in Table~\ref{tab:angles_details} (Appendix~\ref{annex:rotations_details}), we observe a difference with those obtained with the reference rigid registration based on the POIs. 
The wrong rotations of the rigid ICP based on the surfaces come from the truncation of the source which can be interpreted as less constraints for the registration problem.
Similar results are observed for the rigid registration guided by the partial varifold term.
In such cases, the LDDMM fails to compensate for the error that causes the non-rigid deformation to start in a local minimum.
It is illustrated in Fig.~\ref{fig:volume_comparison_04}, where the vessel and the tumor are clearly mismatched for the volumes deformed by approaches using a rigid transformation.  

Inherently, we will not be able to do better than the reference method based on POIs, but we observe that the surface-based registration provides satisfactory results on the whole. When viewing the results, the points of interest are quickly found from one volume to another, even for Patient 14 which is the outlier. Even in this case (first row of Appendix~\ref{annex:worst_case}) we can see that the structures are not so far apart visually, which illustrates a certain robustness of the registrations.
In particular, the partial varifold term allows both rigid and non-rigid consistent registrations with respect to the metric on the POIs.

Fig.~\ref{fig:volume_comparison_56} displays the results for Patient 11. This case corresponds to the best result in terms of POIs distance ($2.77mm$), which is achieved by the rigid method guided by the ICP. Yet this case illustrates that a good alignment of the POIs does not guarantee a good alignment of the tumors: tumors boundaries have been highlighted for better visibility.

\subsubsection{Evaluation on the Tumors}  
\begin{figure*}[!t]
\centering
   \label{fig:s_lesions}%
      \includegraphics[clip, width=0.9\textwidth]{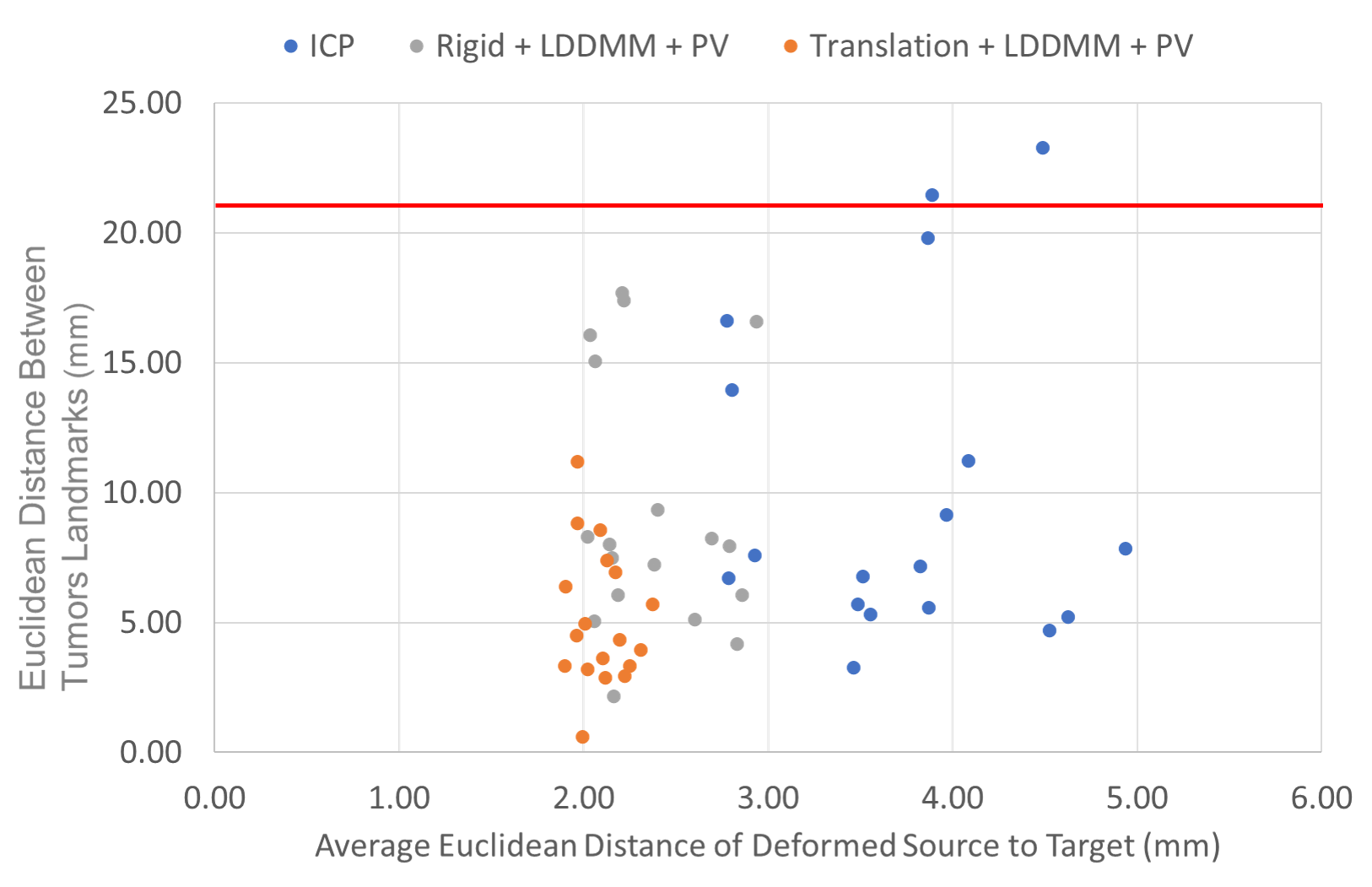}
\caption{\label{fig:scatter_lesions} Registration's evaluation on the Tumors Landmarks. Only the Translation + LDDMM deformation method using Partial Matching preserves the same level of performance on the Tumors Landmarks as for the metric on POIs. The red line corresponds to the average metric on tumors using barycenters registration. The outlier (Patient 14) is excluded from this evaluation.}
\end{figure*}

In the Figure~\ref{fig:scatter_lesions} we show the metric results for the lesions landmarks that we resume per approach in Figure~\ref{fig:results_lesions} and a detailed table is provided in Appendix~\ref{annex:lesions_details}.
The first remarkable result is that the methods using rigid deformations fail to register the tumors closer than about $1cm$ in average while the translation+LDDMM guided by the partial varifold term maintains the same performance level as for the POIs metric with an average distance of $5.13mm$.
In the scatter plot of Fig.~\ref{fig:scatter_lesions} we observe that the rigid ICP and the rigid+LDDMM are more spread than in the scatter plot with the metric on the POIs. 
In particular, the reference deformation optimized with the POIs does not perform well on the tumor registration. The main reason is that none of the POIs is located on the tumor.
In fact the results of rigid ICP and rigid+LDDMM Partial Varifolds are similar to those of the reference rigid registration.
These observations suggest that the rigid deformation, based on surfaces or POIs, is not always the global solution to the volume registration and may lead to local minimum. 
In addition, the non-rigid deformations driven by LDDMM do not improve rigid registration, despite the limitation of the rotation angles. 

\begin{figure*}[!t]
\centering
    \label{fig:box_lesions}
    
      \includegraphics[clip, width=0.9\textwidth]{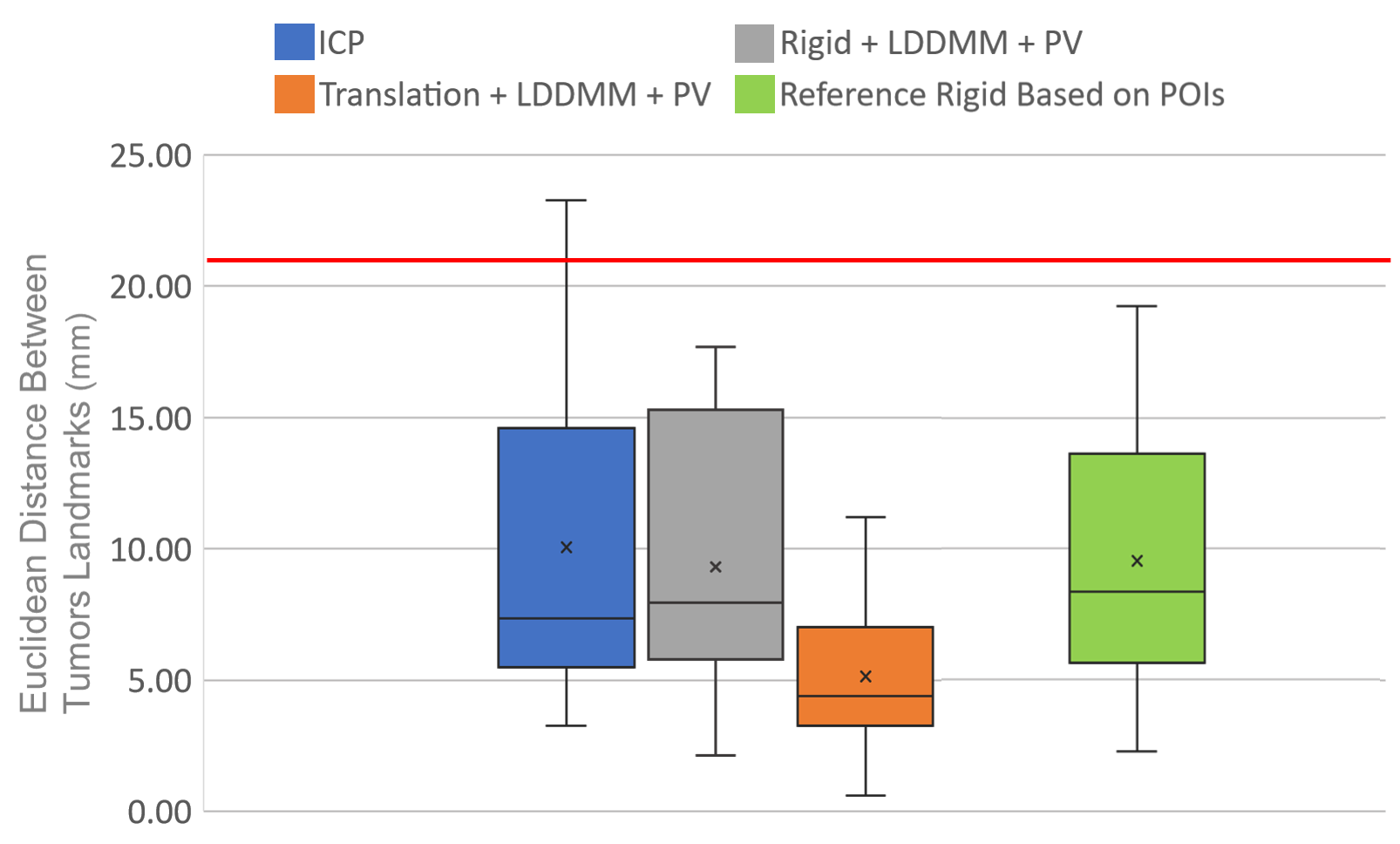}
      
\caption{\label{fig:results_lesions} Registration's evaluation on the tumors landmarks. The rightmost box corresponds to the application of the rigid registration computed with the POIs. The red line corresponds to the average metric on tumors using barycenters registration.}
\end{figure*}
There is a clear difference between these methods and the registration with translation+LDDMM.

Fig.~\ref{fig:volume_comparison_04} (Patient 2), second row, shows a case for which the translation+LDDMM was better overall. 
In such case, the rotations based on the surfaces could not explain the registration between the volumes with a poor result on the lesion metric (about $17.5mm$).
On the contrary, the translation only guided by partial matching shows performance of $10.8mm$ that is further improved by $7mm$ with LDDMM for a distance between the lesion landmarks of $3.62mm$.

In Appendix~\ref{annex:worst_case}, first row, is illustrated the worst case (Patient 14) for which the infiltrated tumor was not annotated, and the liver is hardly visible in CBCT. 
Comparing the results of rigid+LDDMM in Appendix~\ref{annex:lesions_details} and Appendix~\ref{annex:poi_details}, we see that the LDDMM deformations fail to improve significantly the rigid registration guided by the partial matching in the space of varifolds both regarding the POIs metric and the tumors one.
The best initialization for the non-rigid deformations of LDDMM is the translation, providing consistent results for both metrics.

\subsection{Discussion}

We applied the proposed partial matching dissimilarity term to the registration of pre-op CT volumes on CBCT volumes acquired during the interventions based on segmented liver surfaces. 
These surfaces, truncated in CBCT, present non-rigid deformations between them due to differences in time point and patient stances in each modality. 
Partial matching can be used in a rigid registration process, providing results equivalent to a standard method like ICP.
It is important to note that the registered surfaces in this application are relatively smooth, which favors the ICP for the rigid registration part. However, this approach in the non-rigid case can lead to projecting several points on one point of the target, and does not take into account the local orientations of the objects or their resolution. On less regular anatomies, one could expect less good performances. 
Moreover, this approach tends to minimize the average distance from the deformed source to the target. In some cases, as for patient 2 or patient 9, the truncation allows a freedom of deformation which makes the method unsuitable and leads to errors, whereas the reference method generating rigid deformations allowed to obtain good results on both POIs and tumor landmarks.
In some cases (Patients 1, 2, 9 for instance), the reference rigid method is providing better results than the rigid ICP based on the livers surfaces, which ensures that despite an acceptable global rigid deformation, the rigid ICP fails at retrieving the correct one using the surfaces to drive the registrations. 

The proposed Partial Matching can also be used in LDDMM, providing the tools of computational anatomy and allowing non-rigid, yet regular, deformations.
The diffeomorphic deformations that this generates lead to accurate registration of the surfaces of the livers (about $2mm$ on average), which gives the physician a first tool to easily compare CT and CBCT volumes during the procedure. 
However, the areas of interest for the procedures are also within the liver, and care must be taken to ensure that the non-rigid deformations generated from the surfaces generalize well within the liver volume. 

The distance between the Points of Interest, close to the bifurcation of the proper hepatic artery, provide a first evaluation metric to the registration methods. 
However, since they do not cover the volume of the livers correctly, it does not allow to discriminate between the deformation models.
Regarding their location, a rigid registration is sufficient to align them when it fails to extend to the tumors landmarks as illustrated in Fig.~\ref{fig:volume_comparison_56}.
In particular, the extremities of the livers seem to be deformed by non-rigid deformations which can be explained by the better performance of the translation+LDDMM regarding the distance on the tumor landmarks.
In this case, the rigid deformations with rotation lead to a local minimum that the LDDMM are unable to compensate.
On the contrary, translation+LDDMM provide a consistent registration of both POIs and tumors.
This last finding indicates that the correct deformations to generate to register a liver from a CBCT acquisition to the liver from the CT acquisition are translations and local non-rigid deformations without large (or even no) rotations.

Although the methods based on surface registrations are sensitive to surface extraction, we have seen through one case illustrated in Appendix~\ref{annex:worst_case} that the proposed registrations remain suitable, despite the outlier (Patient 14) with respect to the metric on the POIs.
In particular natural the regularization of the LDDMM associated to the one proposed in this paper provide a smooth and consistent non rigid registration of the surface and its ambient space, ensuring realistic registrations of the volumes. 

Regarding the computation time, we are studying a non-rigid multi-modality volume registration solution with partial matching. The current computation time does not allow to use this registration as is in a real time procedure, however there are many levers to accelerate the diffeomorphic deformations, main source of computation cost. These solutions, such as limiting the number of control points (and therefore variables), could allow the use of this solution in an application used during the procedures. 

The purpose of this study is to provide a visualization tool for the transcatheter directed liver therapies in minimally invasive procedures.
The results obtained with the translation+LDDMM method thus provide robust metrics around $5mm$ on average, which is useful for physicians in the perspective of navigating their tools in the patient's anatomy to locate structures that are hardly visible in the CBCT used during the procedure. 
Such registrations could facilitate the physician's intervention by providing, for example through image fusion, an improved visualization of the pre-procedure CT volume tumor placed in the CBCT volume. 
This would allow the physician to avoid redoing a traditional CBCT to see the tumor correctly, thus limiting the X-ray dose sent to the patient and the time of the procedure.

\section{Conclusion}

In this paper we adapted dissimilarity metrics in the space of varifolds to guide partial shape matching and registration. 
This data attachment term is suited to different shapes comparison such as unions of curves or surfaces.
As a clinical application, we showed that this new partial matching term is suitable for the registration of a truncated surface onto a complete one, providing realistic feature-based CT-CBCT volumes registration. 
Regarding the simplicity of the shapes, the rigid ICP shows equivalent results to rigid deformations associated with the proposed partial matching. 
Yet this term can also be associated with a LDDMM registration and allow diffeomorphic deformations.
Provided a correct feature extraction, such registrations could facilitate the physicians procedures during transcatheter directed liver therapies to treat primary and secondary liver malignancies.

In this application we only used one feature extracted from the volumes. In order to better align the CT and CBCT volumes, we could extend the feature extraction to different structures that can be observed in both modalities, such as the hepatic vascular tree, which is partially visible in the CT and in detail in the CBCT.
The dissimilarity term introduced can also be modified to fit the problem of registering a complete shape onto a truncated one. 
The main bottleneck of our approach is the risk of shrinkage when no easy registration is possible. Apart from regularization proposed in this paper, a promising lead to tackle this issue would be to also find a subset of the target to include in the source, such as done in \cite{Bronstein2009}.


\acks{This work was supported by GE Healthcare. We also would like to thank Raphael Doustaly for his technical support and discussions on the subject of multi-modality volumes registration.}

%
\ethics{The work follows appropriate ethical standards in conducting research and writing the manuscript, following all applicable laws and regulations regarding treatment of animals or human subjects.}

\coi{We declare we don't have conflicts of interest.}

\newpage

\appendix\label{Annex}

\section{ POIs Detailed Results per Patients}\label{annex:poi_details}

\begin{table}[!ht]
\centering
\begin{tabular}{|c|c|c|c|c|c||c|}
\hline
Patients & ICP Rigid & Rigid & \begin{tabular}[c]{@{}l@{}}Rigid\\+ LDDMM\end{tabular} & Translation & \begin{tabular}[c]{@{}l@{}}Translation\\+ LDDMM\end{tabular} & \begin{tabular}[c]{@{}l@{}}ICP Rigid\\ Based PoIs\end{tabular}\\ 
\hline \hline 
\centering
Patient 0  & 5.18 & 4.29 & \textbf{4.15} & 18.6 & 4.71 & 2.47\\ 
Patient 1  & \textbf{7.33} & 22.5 & 15.8 & 14.1 & 7.86 & 4.32\\ 
Patient 2  & 9.75 & 9.88 & 8.53 & 12.8 & \textbf{4.94} & 4.92\\ 
Patient 3  & 5.9 & 5.97 & \textbf{5.09} & 9.09 & 6.14 & 2.87\\ 
Patient 4  & \textbf{5.04} & 5.95 & 6.05 & 8.31 & 6.89 & 5.66\\ 
Patient 5  & 5.1 & 6.14 & 4.89 & 5.06 & \textbf{3.37} & 2.26\\ 
Patient 6  & \textbf{4.26} & 4.42 & 6.08 & 6.91 & 10.8 & 1.28\\ 
Patient 7  & 4.06 & \textbf{2.93} & 3.75 & 6.36 & 3.6 & 1.74\\ 
Patient 8  & 4.72 & \textbf{3.55} & 4.76 & 5.87 & 3.67 & 2.36\\ 
Patient 9  & 13.7 & 7.4 & 9.14 & 14.0 & \textbf{6.26} & 1.66\\ 
Patient 10  & 3.7 & \textbf{2.93} & 2.88 & 5.25 & \textbf{2.93} & 1.65\\ 
Patient 11  & \textbf{2.77} & 3.48 & 3.61 & 8.87 & 4.22 & 1.48\\ 
Patient 12  & 5.97 & \textbf{5.31} & 6.02 & 9.03 & 6.3 & 3.02\\ 
Patient 13  & 6.53 & 5.71 & \textbf{5.56} & 13.4 & 5.6 & 2.47\\ 
Patient 14  & 18.9 & 22.3 & 22.8 & 28.1 & \textbf{13.6} & 1.69\\ 
Patient 15  & 2.97 & \textbf{2.3} & 2.55 & 6.46 & 3.59 & 2.64\\ 
Patient 16  & 4.12 & \textbf{3.34} & 3.75 & 4.07 & 4.4 & 2.28\\ 
Patient 17  & 6.17 & 11.0 & \textbf{3.1} & 5.72 & 5.32 & 2.34\\ 
Patient 18  & 7.17 & 7.05 & 5.74 & 13.9 & \textbf{5.76} & 1.89\\ 
\hline
Average     & 6.49 & 7.18 & 6.54 & 10.3 & \textbf{5.79} & 2.58\\ 
Stdev       & 3.93 & 5.83 & 4.95 & 5.9 & \textbf{2.66} & 1.18\\ 
Median      & 5.18 & 5.71 & 5.09 & 8.87 & \textbf{5.32} & 2.34\\ 
\hline
\end{tabular}
\caption{Average distance between the Points of Interest per case.}
\label{tab:poi_details}
\end{table}

\newpage

\section{ Lesions Detailed Results per Patients}\label{annex:lesions_details}

\begin{table}[!ht]
\centering
\begin{tabular}{|c|c|c|c|c|c||c|}
\hline
Patients & ICP Rigid & Rigid & \begin{tabular}[c]{@{}l@{}}Rigid\\+ LDDMM\end{tabular} & Translation & \begin{tabular}[c]{@{}l@{}}Translation\\+ LDDMM\end{tabular} & \begin{tabular}[c]{@{}l@{}}ICP Rigid\\ Based PoIs\end{tabular}\\ 
\hline \hline 
Patient 0  & 5.19 & 3.94 & 5.11 & 17.6 & \textbf{3.31} & 13.6\\ 
Patient 1  & 11.2 & 18.2 & 6.05 & 1.41 & \textbf{2.93} & 5.2\\ 
Patient 2  & 19.8 & 18.8 & 17.4 & 10.8 & \textbf{3.62} & 2.21\\ 
Patient 3  & 16.6 & 16.8 & 16.0 & 16.7 & \textbf{11.2} & 14.0\\ 
Patient 4  & 7.14 & 9.34 & 7.92 & 9.78 & \textbf{6.91} & 13.7\\ 
Patient 5  & 5.68 & 6.77 & 6.03 & 7.25 & \textbf{3.18} & 8.0\\ 
Patient 6  & 4.67 & 4.32 & \textbf{4.15} & 4.37 & 5.7 & 12.0\\ 
Patient 7  & 3.24 & 4.02 & 2.14 & 2.8 & \textbf{0.59} & 6.39\\ 
Patient 8  & 6.76 & 6.17 & 7.48 & 6.77 & \textbf{4.93} & 5.76\\ 
Patient 9  & 23.3 & 18.4 & 16.6 & 6.86 & \textbf{3.94} & 8.68\\ 
Patient 10  & 13.9 & 14.4 & 15.1 & 13.0 & \textbf{8.81} & 9.15\\ 
Patient 11  & 9.12 & 7.49 & 9.32 & 3.66 & \textbf{2.86} & 18.7\\ 
Patient 12  & 7.56 & 8.22 & 8.0 & 6.73 & \textbf{4.48} & 4.04\\ 
Patient 13  & 21.4 & 21.2 & 17.7 & 22.7 & \textbf{8.55} & 13.4\\ 
Patient 14  & None & None & None & None & None & None\\ 
Patient 15  & 6.69 & 7.36 & 8.28 & 9.49 & \textbf{6.36} & 19.3\\ 
Patient 16  & 5.31 & 4.7 & 5.03 & 6.96 & \textbf{3.3} & 6.97\\ 
Patient 17  & \textbf{5.56} & 12.5 & 8.22 & 7.29 & 7.37 & 2.87\\ 
Patient 18  & 7.83 & 7.75 & 7.21 & 12.0 & \textbf{4.32} & 7.44\\ 
\hline
Average    & 10.1 & 10.6 & 9.31 & 9.23 & \textbf{5.13} & 9.52\\  
StDev      & 6.06 & 5.71 & 4.8 & 5.37 & \textbf{2.56} & 5.08\\ 
Median     & 7.35 & 7.98 & 7.96 & 7.27 & \textbf{4.4} & 8.34
\\
\hline
\end{tabular}
\caption{Average distance between the tumors landmarks per case. The invasive tumor could not be annotated for Patient 12. }
\label{tab:lesions_details}
\end{table}

\newpage

\section{Rotations Deformations Comparison}\label{annex:rotations_details}

\begin{table}[ht]
\centering
\begin{tabular}{|c|*3c|*3c|}
 \multicolumn{1}{c}{}  &  \multicolumn{3}{c}{Rigid ICP Based POIs} & \multicolumn{3}{c}{Rigid ICP Based Surfaces}\\
\hline  
Patients & X & Y & Z & X & Y & Z\\ 
\hline \hline 
Patient 0  & -1.15 & -11.52 & 14.2 & -1.53 & -12.18 & 10.7\\ 
Patient 1  & -5.58 & -11.68 & 0.8 & -9.69 & -6.51 & 3.89\\ 
Patient 2  & -8.66 & -6.28 & -0.03 & -17.39 & -2.59 & 0.64\\ 
Patient 3  & -13.49 & -0.11 & 2.58 & -13.03 & 0.51 & 1.33\\ 
Patient 4  & -9.02 & 3.17 & -10.8 & -8.37 & 0.93 & -9.45\\ 
Patient 5  & -2.37 & -4.13 & -2.43 & -4.96 & -1.93 & 1.89\\ 
Patient 6  & -5.09 & -1.27 & -3.64 & -6.42 & 0.65 & -4.96\\ 
Patient 7  & -6.13 & 2.37 & -5.04 & -4.85 & 0.85 & -3.43\\ 
Patient 8  & -6.36 & 4.98 & 6.14 & -6.81 & 3.45 & 2.47\\ 
Patient 9  & -0.77 & -5.45 & 8.57 & -3.49 & -12.56 & 6.15\\ 
Patient 10  & 2.49 & 6.94 & 3.18 & -0.72 & 6.71 & 3.68\\ 
Patient 11  & -10.15 & -5.21 & 1.52 & -9.34 & -3.62 & 0.27\\ 
Patient 12  & 5.1 & 4.96 & -6.6 & 5.3 & 4.18 & -4.14\\ 
Patient 13  & -5.25 & -2.79 & 11.2 & -2.5 & -6.93 & 10.2\\ 
Patient 14  & 3.3 & -12.94 & -4.14 & -9.25 & -16.26 & 6.0\\ 
Patient 15  & -2.38 & 6.39 & -5.4 & -3.88 & 6.68 & -5.35\\ 
Patient 16  & 1.71 & -1.56 & 1.45 & -2.71 & -2.01 & 0.31\\ 
Patient 17  & -0.93 & -1.01 & 4.24 & -4.11 & -2.33 & 7.57\\ 
Patient 18  & -3.28 & 10.3 & -4.61 & 4.76 & 9.8 & 0.54\\ 
\hline
\end{tabular}
\caption{Rigid ICP Rotation Values in degree along each axis, based on the Points of Interest and based on the Surfaces. In bold are highlighted the most important differences in terms of rotation angles.}
\label{tab:angles_details}
\end{table}

\newpage

\section{Worst Case Scenario : Bad Feature Extraction}\label{annex:worst_case}

\begin{figure*}[ht]
\begin{minipage}{.99\linewidth}
\centering
\centering
\begin{minipage}{.31\linewidth}\centering \small Rigid \end{minipage}
\hfill\begin{minipage}{.32\linewidth}\centering \small Rigid+LDDMM \end{minipage}
\hfill\begin{minipage}{.32\linewidth}\centering \small Translation+LDDMM  \end{minipage}
\begin{minipage}{.31\linewidth}\centering \small with ICP \end{minipage}
\hfill\begin{minipage}{.32\linewidth}\centering \small with Partial Varifold \end{minipage}
\hfill\begin{minipage}{.32\linewidth}\centering \small with Partial Varifold \end{minipage}
\vspace{-5mm}
\centering
   \subfloat[POIs metric: 18.9mm, \\ Surface metric : 10.1mm \label{fig:ICP_12}]{%
      \includegraphics[clip, width=0.28\textwidth]{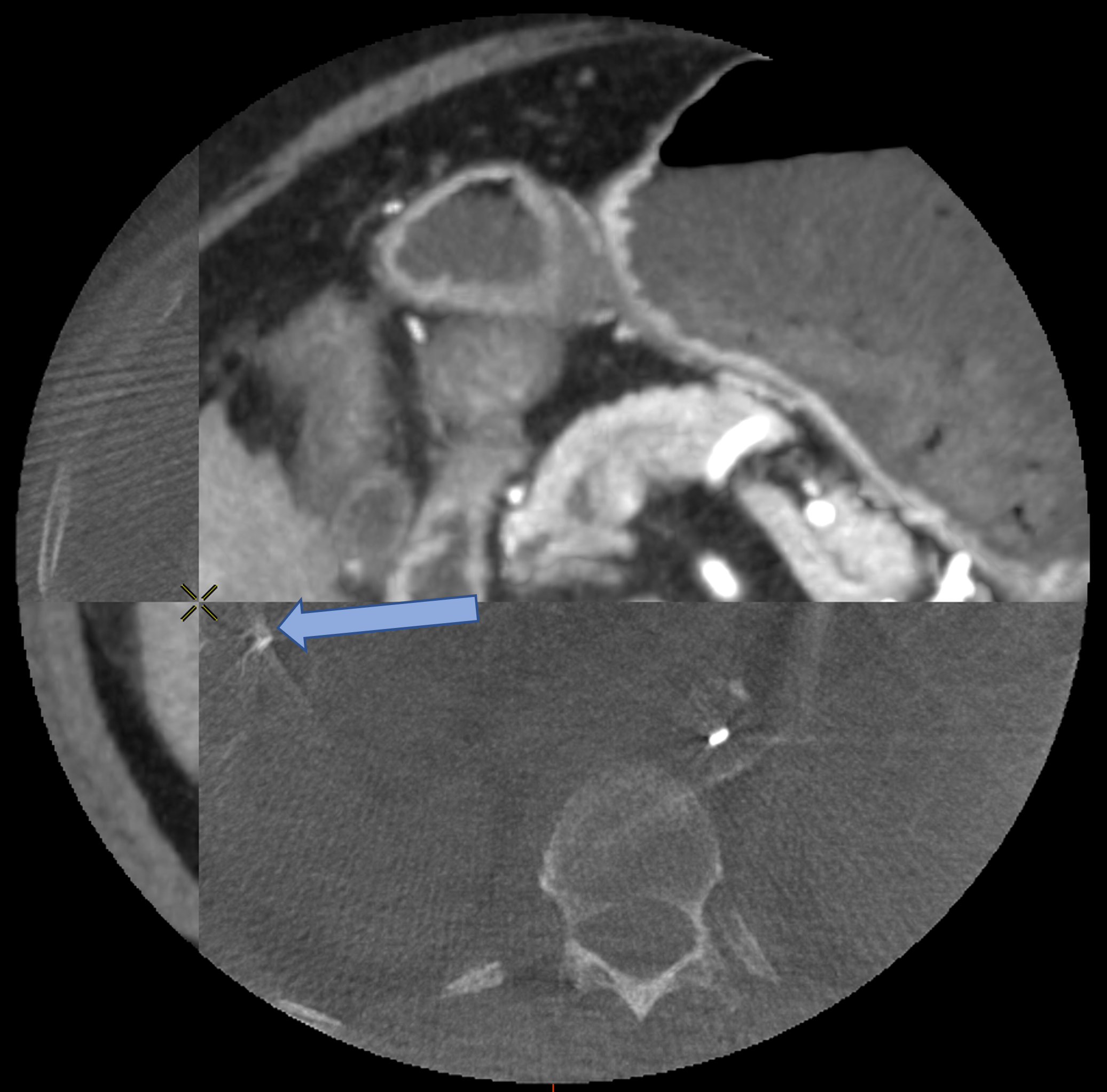}}
\hspace{0.05\textwidth}
\centering
\centering
   \subfloat[POIs metric: 22.8mm, \\ Surface metric : 6.06mm\label{fig:LARAP_12}]{%
      \includegraphics[clip, width=0.28\textwidth]{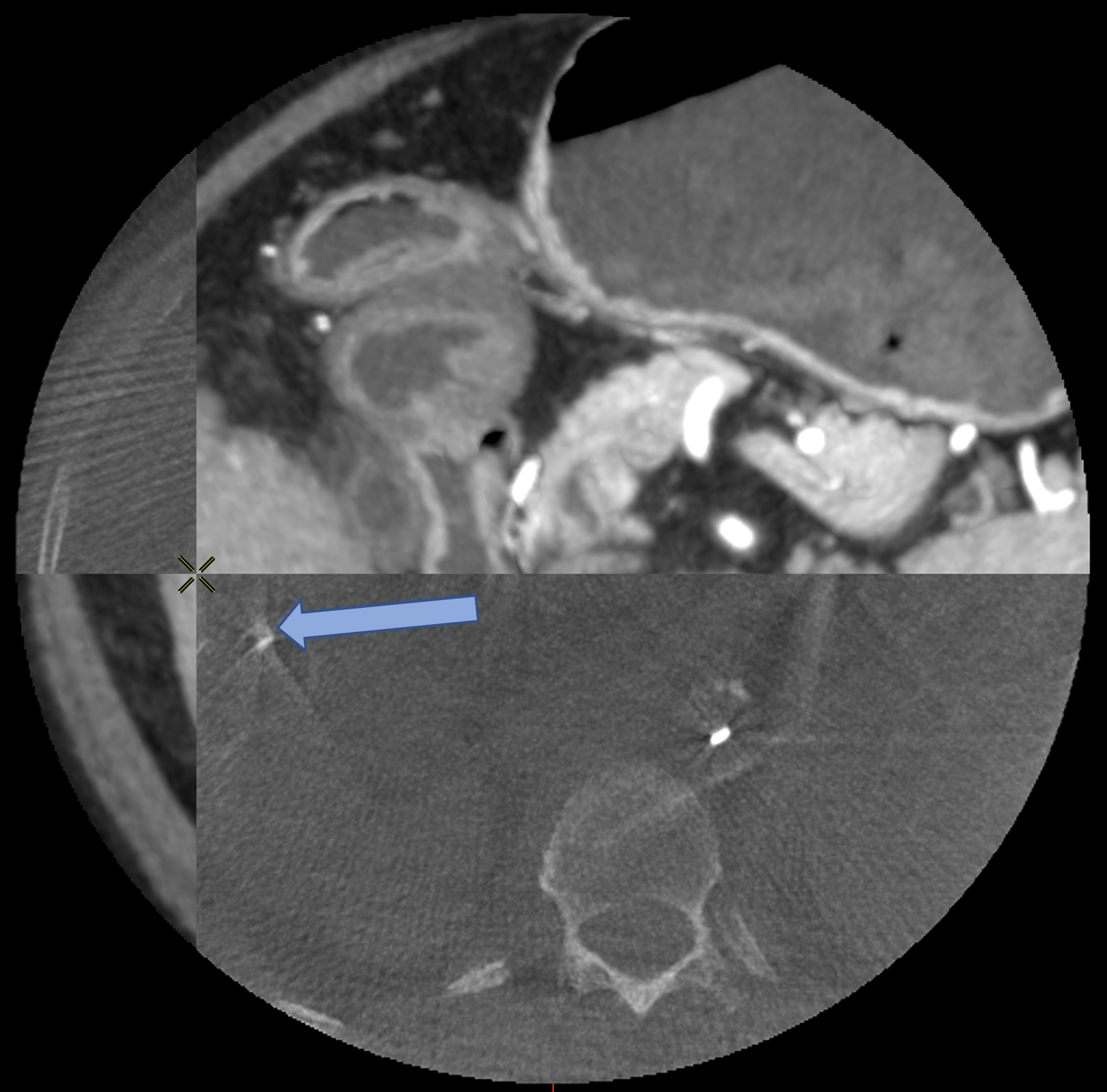}}
\hspace{0.05\textwidth}
\centering
\centering
   \subfloat[POIs metric: 13.5mm, \\ Surface metric : 4.22mm\label{fig:sagi_arap_12}]{%
      \includegraphics[clip, width=0.28\textwidth]{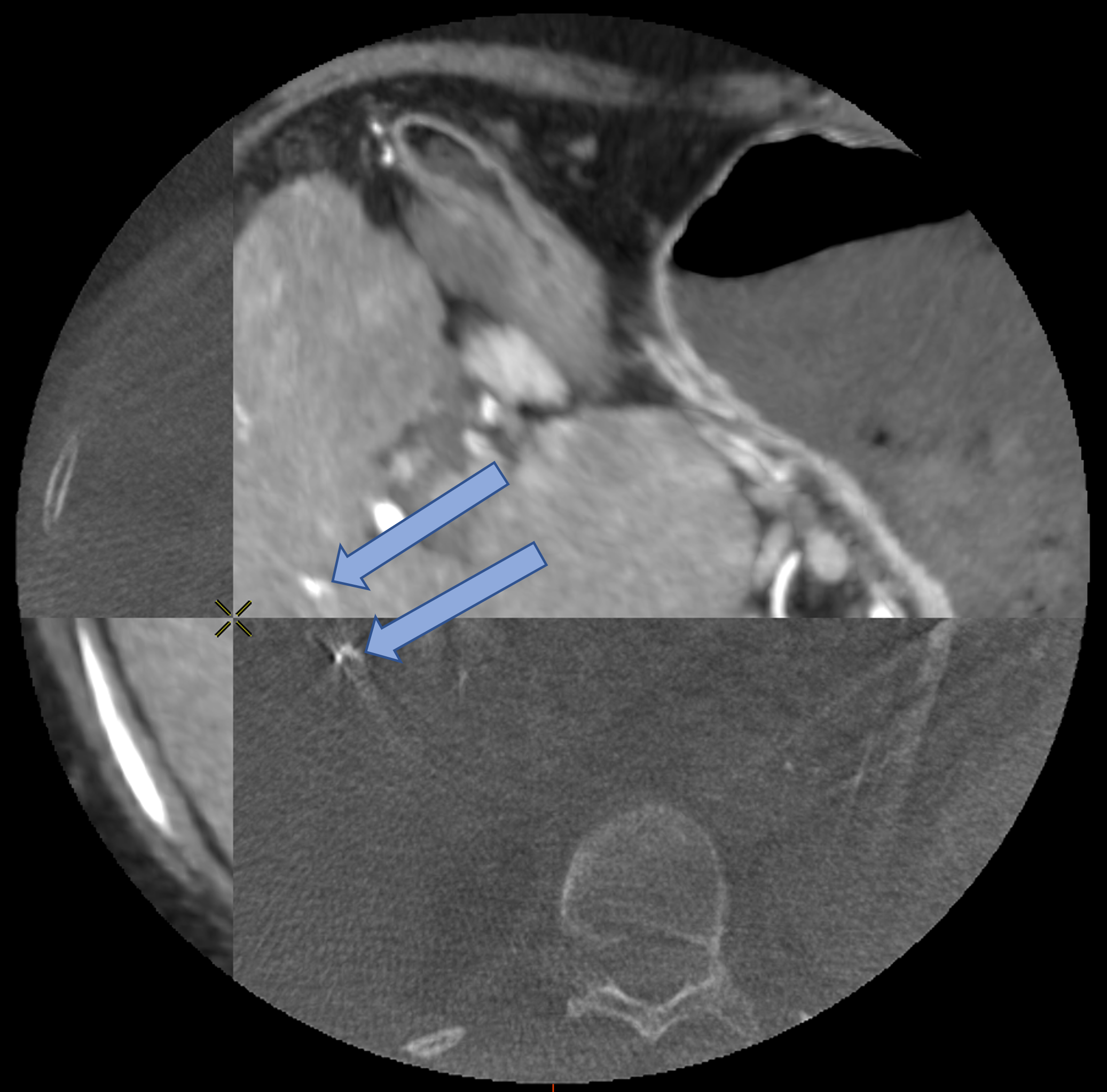}}
\par
\end{minipage}\par\medskip

\caption{\label{fig:volume_comparison_12} Tiled visualization of the registrations for Patient 14. The infiltrated tumor could not be annotated. The tiles of the CBCT target volume are the dark ones, those of the deformed CT volumes are the light ones. }
\end{figure*}


\begin{thebibliography}{37}
\providecommand{\natexlab}[1]{#1}
\providecommand{\url}[1]{\texttt{#1}}
\expandafter\ifx\csname urlstyle\endcsname\relax
  \providecommand{\doi}[1]{doi: #1}\else
  \providecommand{\doi}{doi: \begingroup \urlstyle{rm}\Url}\fi

\bibitem[Aiger et~al.(2008)Aiger, Mitra, and Cohen-Or]{Aiger2008}
Dror Aiger, Niloy Mitra, and Daniel Cohen-Or.
\newblock 4-points congruent sets for robust pairwise surface registration.
\newblock \emph{35th International Conference on Computer Graphics and
  Interactive Techniques (SIGGRAPH'08)}, 27, 08 2008.
\newblock \doi{10.1145/1399504.1360684}.

\bibitem[Antonsanti et~al.(2021)Antonsanti, Glaun\`es, Benseghir, Jugnon, and
  Kaltenmark]{Antonsanti2021}
Pierre-Louis Antonsanti, Joan Glaun\`es, Thomas Benseghir, Vincent Jugnon, and
  Ir\`ene Kaltenmark.
\newblock Partial matching in the space of varifolds.
\newblock \emph{Information Processing in Medical Imaging}, 2021.

\bibitem[Bashiri et~al.(2018)Bashiri, Baghaie, Rostami, Yu, and
  D’Souza]{Bashiri2018}
Fereshteh~S. Bashiri, Ahmadreza Baghaie, Reihaneh Rostami, Zeyun Yu, and Roshan
  D’Souza.
\newblock Multi-modal medical image registration with full or partial data: A
  manifold learning approach.
\newblock \emph{Journal of Imaging}, 5:\penalty0 5, 12 2018.
\newblock \doi{10.3390/jimaging5010005}.

\bibitem[Bauer et~al.(2021)Bauer, Russ, Waldkirch, Tönnes, Segars, Schad,
  Zöllner, and Golla]{Bauer2021}
Dominik Bauer, Tom Russ, Barbara Waldkirch, Christian Tönnes, William Segars,
  Lothar Schad, Frank Zöllner, and Alena-Kathrin Golla.
\newblock Generation of annotated multimodal ground truth datasets for
  abdominal medical image registration.
\newblock \emph{International Journal of Computer Assisted Radiology and
  Surgery}, 16, 05 2021.
\newblock \doi{10.1007/s11548-021-02372-7}.

\bibitem[Beg et~al.(2005)Beg, Miller, Trouv\'e, and Younes]{Beg2005}
Mirza~Faisal Beg, Michael Miller, Alain Trouv\'e, and Laurent Younes.
\newblock Computing large deformation metric mappings via geodesic flows of
  diffeomorphisms.
\newblock \emph{International Journal of Computer Vision}, 61:\penalty0
  139--157, 02 2005.
\newblock \doi{10.1023/B:VISI.0000043755.93987.aa}.

\bibitem[Bronstein et~al.(2009)Bronstein, Bronstein, Bruckstein, and
  Kimmel]{Bronstein2009}
Alexander Bronstein, Michael Bronstein, Alfred Bruckstein, and Ron Kimmel.
\newblock Partial similarity of objects, or how to compare a centaur to a
  horse.
\newblock \emph{International Journal of Computer Vision}, 84:\penalty0
  163--183, 08 2009.
\newblock \doi{10.1007/s11263-008-0147-3}.

\bibitem[Bronstein et~al.(2006)Bronstein, Bronstein, and Kimmel]{Bronstein2006}
Alexander~M. Bronstein, Michael~M. Bronstein, and Ron Kimmel.
\newblock Generalized multidimensional scaling: A framework for
  isometry-invariant partial surface matching.
\newblock \emph{Proceedings of the National Academy of Sciences}, 103\penalty0
  (5):\penalty0 1168--1172, 2006.
\newblock ISSN 0027-8424.
\newblock \doi{10.1073/pnas.0508601103}.
\newblock URL \url{https://www.pnas.org/content/103/5/1168}.

\bibitem[Charlier et~al.(2021)Charlier, Feydy, Glaun\`es, Collin, and
  Durif]{JMLR:v22:20-275}
Benjamin Charlier, Jean Feydy, Joan~Alexis Glaun\`es, François-David Collin,
  and Ghislain Durif.
\newblock Kernel operations on the gpu, with autodiff, without memory
  overflows.
\newblock \emph{Journal of Machine Learning Research}, 22\penalty0
  (74):\penalty0 1--6, 2021.
\newblock URL \url{http://jmlr.org/papers/v22/20-275.html}.

\bibitem[Charon and Trouv\'e(2013)]{Charon2013}
Nicolas Charon and Alain Trouv\'e.
\newblock The varifold representation of nonoriented shapes for diffeomorphic
  registration.
\newblock \emph{SIAM Journal on Imaging Sciences}, 6\penalty0 (4):\penalty0
  2547--2580, 2013.
\newblock \doi{10.1137/130918885}.

\bibitem[Charon et~al.(2020)Charon, Charlier, Glaun\`es, Gori, and
  Roussillon]{Charon2020}
Nicolas Charon, Benjamin Charlier, Joan Glaun\`es, Pietro Gori, and Pierre
  Roussillon.
\newblock 12 - fidelity metrics between curves and surfaces: currents,
  varifolds, and normal cycles.
\newblock In Xavier Pennec, Stefan Sommer, and Tom Fletcher, editors,
  \emph{Riemannian Geometric Statistics in Medical Image Analysis}, pages 441
  -- 477. Academic Press, 2020.
\newblock ISBN 978-0-12-814725-2.
\newblock \doi{https://doi.org/10.1016/B978-0-12-814725-2.00021-2}.
\newblock URL
  \url{http://www.sciencedirect.com/science/article/pii/B9780128147252000212}.

\bibitem[Christensen et~al.(1996)Christensen, Rabbitt, and
  Miller]{Christensen1996}
G.E. Christensen, R.D. Rabbitt, and M.I. Miller.
\newblock Deformable templates using large deformation kinematics.
\newblock \emph{IEEE Transactions on Image Processing}, 5\penalty0
  (10):\penalty0 1435--1447, 1996.
\newblock \doi{10.1109/83.536892}.

\bibitem[Feragen et~al.(2015)Feragen, Petersen, de~Bruijne, and
  et~al.]{Feragen2015}
Aasa Feragen, Jens Petersen, Marleen de~Bruijne, and et~al.
\newblock Geodesic atlas-based labeling of anatomical trees: Application and
  evaluation on airways extracted from ct.
\newblock \emph{IEEE Transactions on Medical Imaging}, 34:\penalty0 1212--1226,
  2015.

\bibitem[Feydy et~al.(2017)Feydy, Charlier, Vialard, and Peyr\'e]{Feydy2017}
Jean Feydy, Benjamin Charlier, François-Xavier Vialard, and Gabriel Peyr\'e.
\newblock Optimal transport for diffeomorphic registration.
\newblock \emph{Lecture Notes in Computer Science}, page 291–299, 2017.
\newblock ISSN 1611-3349.
\newblock URL \url{http://dx.doi.org/10.1007/978-3-319-66182-7_34}.

\bibitem[Ghosn et~al.(2021)Ghosn, Derbel, andN. Oubaya, Mul\'e, Chalaye,
  Regnault, Amaddeo, Itti, Luciani, Kobeiter, and Tacher]{Ghosn2020}
Mario Ghosn, H.~Derbel, R.~Kharrat andN. Oubaya, S.~Mul\'e, J.~Chalaye,
  H.~Regnault, G.~Amaddeo, E.~Itti, A.~Luciani, H.~Kobeiter, and V.~Tacher.
\newblock Prediction of overall survival in patients with hepatocellular
  carcinoma treated with y-90 radioembolization by imaging response criteria.
\newblock \emph{Diagnostic and Interventional Imaging}, 102:\penalty0 35--44,
  08 2021.

\bibitem[Glaun\`es(2005)]{Glaunes2005}
Joan Glaun\`es.
\newblock \emph{Transport par diff\'eomorphismes de points, demesures et de
  courants pour la comparaison de formes et l’anatomie num\'erique.}
\newblock PhD thesis, Universit\'e Paris 13, 2005.

\bibitem[{Halimi} et~al.(2019){Halimi}, {Litany}, {Rodol\`a}, {Bronstein}, and
  {Kimmel}]{Halimi2019}
O.~{Halimi}, O.~{Litany}, E.~R. {Rodol\`a}, A.~M. {Bronstein}, and R.~{Kimmel}.
\newblock Unsupervised learning of dense shape correspondence.
\newblock \emph{2019 IEEE/CVF Conference on Computer Vision and Pattern
  Recognition (CVPR)}, pages 4365--4374, 2019.
\newblock \doi{10.1109/CVPR.2019.00450}.

\bibitem[Halimi et~al.(2020)Halimi, Imanuel, Litany, Trappolini, Rodol{\`{a}},
  Guibas, and Kimmel]{Halimi2020}
Oshri Halimi, Ido Imanuel, Or~Litany, Giovanni Trappolini, Emanuele
  Rodol{\`{a}}, Leonidas~J. Guibas, and Ron Kimmel.
\newblock The whole is greater than the sum of its nonrigid parts.
\newblock \emph{CoRR}, abs/2001.09650, 2020.
\newblock URL \url{https://arxiv.org/abs/2001.09650}.

\bibitem[Hsieh and Charon(2021)]{Hsieh2021}
Hsi-Wei Hsieh and Nicolas Charon.
\newblock Diffeomorphic registration with density changes for the analysis of
  imbalanced shapes.
\newblock \emph{Information Processing in Medical Imaging}, pages 31--42, 2021.

\bibitem[Jiang et~al.(2021)Jiang, Ma, Xiao, Shao, and Guo]{Jiang2021}
Xingyu Jiang, Jiayi Ma, Guobao Xiao, Zhenfeng Shao, and Xiaojie Guo.
\newblock A review of multimodal image matching: Methods and applications.
\newblock \emph{Information Fusion}, 73:\penalty0 22--71, 2021.
\newblock ISSN 1566-2535.
\newblock \doi{https://doi.org/10.1016/j.inffus.2021.02.012}.
\newblock URL
  \url{https://www.sciencedirect.com/science/article/pii/S156625352100035X}.

\bibitem[Kaltenmark and Trouv{\'e}(2018)]{Kaltenmark2018}
Ir{\`e}ne Kaltenmark and Alain Trouv{\'e}.
\newblock {Estimation of a Growth Development with Partial Diffeomorphic
  Mappings}.
\newblock \emph{Quaterly of Applied MAthematics}, 77:\penalty0 227--267,
  November 2018.

\bibitem[Kaltenmark et~al.(2017)Kaltenmark, Charlier, and
  Charon]{Kaltenmark2017}
Irene Kaltenmark, Benjamin Charlier, and Nicolas Charon.
\newblock A general framework for curve and surface comparison and registration
  with oriented varifolds.
\newblock \emph{Proceedings of the IEEE Conference on Computer Vision and
  Pattern Recognition (CVPR)}, July 2017.

\bibitem[Miller et~al.(2006)Miller, Trouv{\'e}, and Younes]{Miller06}
Michael~I. Miller, Alain Trouv{\'e}, and Laurent Younes.
\newblock Geodesic shooting for computational anatomy.
\newblock \emph{Journal of Mathematical Imaging and Vision}, 24\penalty0
  (2):\penalty0 209--228, Mar 2006.
\newblock ISSN 1573-7683.

\bibitem[Milletari et~al.(2016)Milletari, Navab, and Ahmadi]{Milletari2016}
Fausto Milletari, Nassir Navab, and Seyed-Ahmad Ahmadi.
\newblock V-net: Fully convolutional neural networks for volumetric medical
  image segmentation.
\newblock \emph{2016 fourth international conference on 3D vision (3DV)}, pages
  565--571, 10 2016.
\newblock \doi{10.1109/3DV.2016.79}.

\bibitem[Mori et~al.(2013)Mori, Sakuma, Sato, Barillot, and
  Navab]{Benseghir2013}
Kensaku Mori, Ichiro Sakuma, Yoshinobu Sato, Christian Barillot, and Nassir
  Navab, editors.
\newblock \emph{Iterative Closest Curve: A Framework for Curvilinear Structure
  Registration Application to 2D/3D Coronary Arteries Registration}, Berlin,
  Heidelberg, 2013. Springer Berlin Heidelberg.

\bibitem[Musy et~al.(2021)Musy, Jacquenot, Dalmasso, neoglez, de~Bruin,
  Pollack, Claudi, Badger, icemtel, Sullivan, Lerner, Hrisca, Volpatto,
  Schlömer, RichardScottOZ, Zhou, and ilorevilo]{Musy2021_vedo}
Marco Musy, Guillaume Jacquenot, Giovanni Dalmasso, neoglez, Ruben de~Bruin,
  Ahinoam Pollack, Federico Claudi, Codacy Badger, icemtel, Bane Sullivan,
  Brian Lerner, Daniel Hrisca, Diego Volpatto, Nico Schlömer, RichardScottOZ,
  Zhi-Qiang Zhou, and ilorevilo.
\newblock marcomusy/vedo: 2021.0.7, November 2021.
\newblock URL \url{https://doi.org/10.5281/zenodo.5655358}.

\bibitem[Ovsjanikov et~al.(2012)Ovsjanikov, Ben-Chen, Solomon, Butscher, and
  Guibas]{Ovsjanikov2013}
Maks Ovsjanikov, Mirela Ben-Chen, Justin Solomon, Adrian Butscher, and Leonidas
  Guibas.
\newblock Functional maps: A flexible representation of maps between shapes.
\newblock \emph{ACM Trans. Graph.}, 31\penalty0 (4), July 2012.
\newblock ISSN 0730-0301.
\newblock \doi{10.1145/2185520.2185526}.

\bibitem[Paszke et~al.(2017)Paszke, Gross, Chintala, Chanan, Yang, DeVito, Lin,
  Desmaison, Antiga, and Lerer]{pytorch}
Adam Paszke, Sam Gross, Soumith Chintala, Gregory Chanan, Edward Yang, Zachary
  DeVito, Zeming Lin, Alban Desmaison, Luca Antiga, and Adam Lerer.
\newblock Automatic differentiation in {PyTorch}.
\newblock In \emph{NIPS Autodiff Workshop}, 2017.

\bibitem[Rajagopal and Venkatesan(2016)]{Rajagopal2016}
M.~Rajagopal and AM. Venkatesan.
\newblock Image fusion and navigation platforms for percutaneous image-guided
  interventions.
\newblock \emph{Abdom Radiol (NY)}, 41:\penalty0 620--628, 04 2016.
\newblock \doi{10.1007/s00261-016-0645-7}.

\bibitem[Rodol\`a et~al.(2017)Rodol\`a, Cosmo, Bronstein, Torsello, and
  Cremers]{Rodola2017}
E.~Rodol\`a, L.~Cosmo, M.~M. Bronstein, A.~Torsello, and D.~Cremers.
\newblock Partial functional correspondence.
\newblock \emph{Computer Graphics Forum}, 36\penalty0 (1):\penalty0 222--236,
  2017.
\newblock \doi{https://doi.org/10.1111/cgf.12797}.

\bibitem[Rodol\`a et~al.(2013)Rodol\`a, Albarelli, Bergamasco, and
  Torsello]{Rodola2013}
Emanuele Rodol\`a, Andrea Albarelli, Filippo Bergamasco, and Andrea Torsello.
\newblock A scale independent selection process for 3d object recognition in
  cluttered scenes.
\newblock \emph{International Journal of Computer Vision}, 102, 03 2013.
\newblock \doi{10.1007/s11263-012-0568-x}.

\bibitem[{Sotiras} et~al.(2013){Sotiras}, {Davatzikos}, and
  {Paragios}]{Sotiras2013}
A.~{Sotiras}, C.~{Davatzikos}, and N.~{Paragios}.
\newblock Deformable medical image registration: A survey.
\newblock \emph{IEEE Transactions on Medical Imaging}, 32\penalty0
  (7):\penalty0 1153--1190, July 2013.
\newblock ISSN 1558-254X.
\newblock \doi{10.1109/TMI.2013.2265603}.

\bibitem[Sukurdeep et~al.(2021)Sukurdeep, Bauer, and Charon]{Sukurdeep2021}
Yashil Sukurdeep, Martin Bauer, and Nicolas Charon.
\newblock A new variational model for the analysis of shape graphs with partial
  matching constraints.
\newblock \emph{arXiv preprint arXiv:2105.00678}, 2021.

\bibitem[Tacher et~al.(2015)Tacher, Radaelli, Lin, and Geschwind]{Tacher2015}
V.~Tacher, A.~Radaelli, M.~Lin, and JF. Geschwind.
\newblock How i do it: Cone-beam ct during transarterial chemoembolization for
  liver cancer.
\newblock \emph{Radiology}, 274:\penalty0 320--334, 02 2015.
\newblock \doi{10.1148/radiol.14131925}.

\bibitem[Trouv{\'e}(1995)]{Trouve1995}
Alain Trouv{\'e}.
\newblock Infinite-dimensional group action and pattern-recognition.
\newblock \emph{COMPTES RENDUS DE L ACADEMIE DES SCIENCES SERIE
  I-MATHEMATIQUE}, 321\penalty0 (8):\penalty0 1031--1034, 1995.

\bibitem[van Kaick et~al.(2011)van Kaick, Zhang, Hamarneh, and
  Cohen-Or]{Kaick2011}
Oliver van Kaick, Hao Zhang, Ghassan Hamarneh, and Daniel Cohen-Or.
\newblock A survey on shape correspondence.
\newblock \emph{Computer Graphics Forum}, 30\penalty0 (6):\penalty0 1681--1707,
  2011.
\newblock URL
  \url{https://onlinelibrary.wiley.com/doi/abs/10.1111/j.1467-8659.2011.01884.x}.

\bibitem[van Kaick et~al.(2013)van Kaick, Zhang, and Hamarneh]{Kaick2013}
Oliver van Kaick, Hao Zhang, and Ghassan Hamarneh.
\newblock Bilateral maps for partial matching.
\newblock \emph{Computer Graphics Forum (CGF)}, 09 2013.
\newblock \doi{10.1111/cgf.12084}.

\bibitem[Zhao et~al.(2015)Zhao, Price, Pizer, Niethammer, Alterovitz, and
  Rosenman]{Zhao2015}
Qingyu Zhao, James Price, S.~Pizer, M.~Niethammer, R.~Alterovitz, and
  J.~Rosenman.
\newblock Surface registration in the presence of missing patches and topology
  change.
\newblock In \emph{MIUA}, 2015.

\end{thebibliography}
\end{document}